\documentclass[10pt, preprint]{elsarticle}

\usepackage{amsmath,amssymb}
\usepackage{epsfig}
\usepackage{graphicx}

\usepackage{algorithm}
\usepackage{algorithmic}

\usepackage[usenames]{color}
\definecolor{mygray}{gray}{0.5}

\usepackage{ulem}
\usepackage{xspace} 

\usepackage{subfigure}

\def\herm{\mathrm{H}}

\newcommand{\mps}{{\sc{mps}}\xspace}
\newcommand{\MPS}{{\sc{mps}}\xspace}

\newcommand{\MPO}{{\sc{mpo}}\xspace}
\newcommand{\MPDO}{{\sc{mpdo}}\xspace}

\newcommand{\peps}{{\sc{peps}}\xspace}
\newcommand{\PEPS}{{\sc{peps}}\xspace}

\newcommand{\als}{{\sc{als}}\xspace}
\newcommand{\ALS}{{\sc{als}}\xspace}

\newcommand{\TT}{{\sc{tt}}\xspace}

\newcommand{\parafac}{{\sc{parafac}}\xspace}
\newcommand{\PARAFAC}{{\sc{parafac}}\xspace}
\newcommand{\ParaFac}{{\sc{parafac}}\xspace}

\newcommand{\TTS}{{\sc{tts}}\xspace}

\newcommand{\dmrg}{{\sc{dmrg}}\xspace}

\def\defini#1{\emph{#1}}

\renewcommand{\vec}[1]{\boldsymbol{#1}}

\newcommand{\trace}{\operatorname{tr}}

\newcommand{\Herm}{\mathnormal{{\color{black}\mathrm{H}}}}

\newcommand{\myref}[1]{(\ref{#1})}

\providecommand{\mat}[1]{\begin{pmatrix} #1 \end{pmatrix}}

\begin{document}
\begin{frontmatter}

  \title{Computations in Quantum Tensor Networks}

  \author[sccs]{T. Huckle \corref{corauth}}
  \ead{huckle@in.tum.de}
  \author[sccs]{K.\ Waldherr}
  \ead{waldherr@in.tum.de}
  \author[oc2]{T. Schulte-Herbr{\"u}ggen}
  \ead{tosh@tum.de}
  \cortext[corauth]{Corresponding author}
  \address[sccs]{Technische Universit\"at M\"unchen, Boltzmannstr. 3, 85748 Garching, Germany}
  \address[oc2]{Technische Universit\"at M\"unchen, Lichtenbergstr. 4, 85748 Garching, Germany}

\begin{abstract}
The computation of the ground state (i.e. the eigenvector related to the smallest eigenvalue)
is an important task in the simulation of quantum many-body systems.
As the dimension of the underlying vector space grows exponentially in the number of particles,
one has to consider appropriate subsets promising both convenient approximation properties and
efficient computations.
The variational ansatz for this numerical approach leads to the minimization of the Rayleigh quotient.
The Alternating Least Squares technique is then applied to break down the eigenvector computation to problems
of appropriate size, which can be solved by classical methods.
Efficient computations require fast computation of the matrix-vector product and of the inner product
of two decomposed vectors.
To this end, both appropriate representations of vectors and efficient contraction schemes are
needed.

Here approaches from many-body quantum physics for one-dimensional and two-dimensional systems
(Matrix Product States and Projected Entangled Pair States)
are treated mathematically in terms of tensors.
We give the definition of these concepts,
bring some results concerning uniqueness and numerical stability
and show how computations can be executed efficiently within these concepts.
Based on this overview we present some modifications and generalizations of these concepts
and show that they still allow efficient computations such as applicable contraction schemes.
In this context we consider the minimization of the Rayleigh quotient in terms of the \ParaFac (CP) formalism,
where we also allow different tensor partitions.
This approach makes use of efficient contraction schemes for the calculation of inner products
in a way that can easily be extended to the \MPS format but also to higher dimensional problems.
\end{abstract}


\end{frontmatter}

\pagestyle{myheadings}
\thispagestyle{plain}


\section{Introduction}
\label{seq:intro}
Computations with tensors are getting increasingly important in high dimensional problems.
In particular in quantum many-body physics, a typical problem amounts to finding an accurate
approximation to the smallest eigenvalue (i.e., the ground-state energy) of a hermitian matrix
(representing the Hamiltonian) that is larger than one can store even on a powerful computer.
To this end, in quantum physics techniques like Matrix Product States (\MPS)
or Projected Entangled Pair States (\PEPS) have been developed for representing vectors,
{\em viz.}~eigenstates of quantum systems {\em efficiently}.
In mathematics, besides the Tucker decomposition and the canonical decomposition,
concepts like Tensor Trains (\TT) were introduced.
The examples of \MPS or \peps (in physics) and \TT (in mathematics) express a common interest in powerful numerical
methods specifically designed for coping with high-dimensional tensor networks. ---
Unifying variational approaches to ground-state calculations~\cite{Eisert07}
in a {\em common framework of tensor approximations} will be highly useful, in
particular in view of optimizing numerical algorithms \cite{ALPS,GR06,PMCV10,VCM09}.
Here it is the goal to cast some of the recent developments in mathematical physics
into such a common frame expressed in the terminology of multilinear algebra.
Moreover we present numerical results on the Ising-type Hamiltonian underscoring
the wealth and the potential of such an approach.

In this paper, we address one-dimensional and two-dimensional methods
in a unified frame related to \mps and \peps.
We will introduce some generalization of \mps and the canonical decomposition.
Furthermore, we give a short description of tensor-decomposition methods for $2$D problems.

\subsection*{Scope and Organization}
The paper is organized as follows:
Sections \ref{sec:physicalModelSystem}, \ref{sec:states} and \ref{sec:tensor}
contain an overview of already-known concepts and describe them in the multilinear algebra language:
in Section \ref{sec:physicalModelSystem} we introduce the physical background of the problem setting
and define the matrices involved,
in Section \ref{sec:states} we present representation schemes for states in physically
motivated 1D and 2D arrays and we show how computations can be performed efficiently and
Section \ref{sec:tensor} finally fixes some basics and notations for tensors and tensor-decomposition schemes.

In Section \ref{sec:generalization} we present new ideas of how to generalize these basic concepts, how to
execute calculations efficiently and how to apply them to the ground-state approximation problem.
First numerical results will show the benefit of these newly developed concepts.

\section{Physical Model Systems}
\label{sec:physicalModelSystem}
Consider vectors~$\vec x$ in a complex Hilbert space~$\mathcal H$
representing states of (pure) quantum systems.
The differential equation $\vec{\dot x} = -i H \vec{x}$ (Schr{\"o}dinger's equation of motion)
then governs quantum dynamics (neglecting relaxation)
with the Hamiltonian $H$ being the generator of unitary time evolution.
The Hamiltonian captures the energies of the constituent subsystems (e.g.~spins) as well as
the interaction energies between coupled subsystems.

\begin{figure}[ht]
\centering
\subfigure[1D system of $5$ spins with nearest-neighbor interaction and periodic boundary conditions.]{
\includegraphics[width=0.95\textwidth]{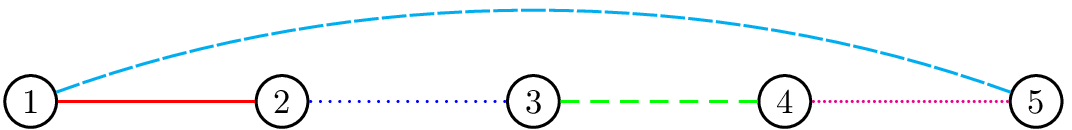}
\label{fig:linspin1}
}
\vfill
\subfigure[Representation of the Hamiltonian related to the physical system illustrated by Figure \ref{fig:linspin1}.]{
\includegraphics[width=0.5\textwidth]{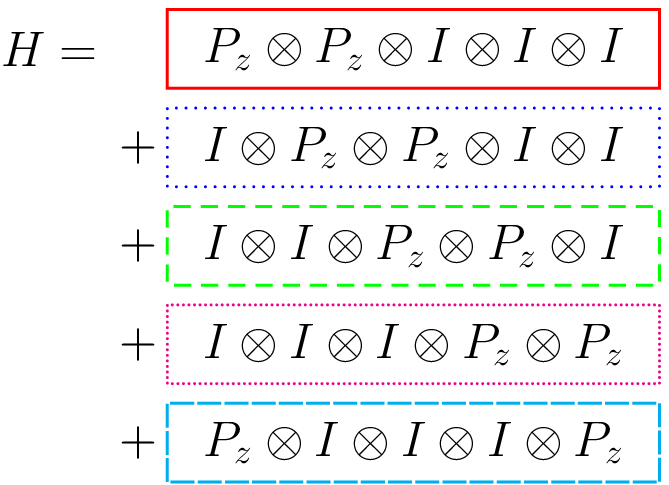}
\label{fig:linspin2}
}
\caption{Example of a linear physical system (a) and the related Hamiltonian (b).}
\end{figure}

For instance, a linear chain of five spins coupled by nearest neighbor interactions
can be depicted as in Figure~\ref{fig:linspin1}.
In the case of \defini{open boundary conditions} (OBC) there is no coupling interaction between particle $1$ and $5$,
while for \defini{periodic boundary conditions} (PBC) there is a non vanishing coupling interaction between $1$ and $5$.

\subsection{Hamiltonian Representations}\label{subsec:HamiltonianRepresentations}
For spin $\tfrac{1}{2}$ particles such as electrons or protons, the spin angular momentum operator
describing their internal degree of freedom (i.e. spin-up and spin-down)
is usually expressed in terms of the \defini{Pauli matrices}
\begin{equation}\label{eq:Pauli}
P_x= \mat{0 & 1 \\ 1 & 0 }, \
P_y= \mat{0 & -i \\ i & 0 } \ \text{and} \
P_z= \mat{1 & 0 \\ 0 & -1}\quad.
\end{equation}
Being traceless and Hermitian, $\{P_x, P_y,P_z\}$ forms a basis of the Lie algebra $\mathfrak{su}(2)$,
while by appending the identity matrix $I$ one obtains a basis of the Lie algebra $\mathfrak{u}(2)$.

Now, spin Hamiltonians are built by summing $M$ terms,
each of them representing a physical (inter)action.
These terms are themselves tensor products of Pauli matrices or identities
\begin{equation}\label{eq:Hamiltonian}
H=\sum_{k=1}^M\underbrace{\alpha_k (Q_{1}^{(k)}\otimes Q_{2}^{(k)}\otimes %
		\cdots \otimes Q_{p}^{(k)} )}_{=:H^{(k)}}= \sum_{k=1}^M H^{(k)}\;,
\end{equation}
where $Q_j^{(k)}$ can be $P_x$, $P_y$, $P_z$ or $I$.

In each summand $H^{(k)}$ most of the $Q_{j}^{(k)}$ are $I$:
{\em local terms} have just one nontrivial tensor factor, while {\em pair interactions} have two of them.
Higher \mbox{$m$-body} interactions (with $m>2$) usually do not occur as physical primitives, but could be
represented likewise by $m$ Pauli matrices in the tensor product representing the $m$-order interaction
term\footnote{For further details, a reader wishing to
approach quantum physics from linear and multilinear algebra may refer to \cite{nakahara2008quantum}.}.

For instance, in the \defini{Ising} ($ZZ$) model \cite{P70} for the 1D chain with $p$ spins
and open boundary conditions, the spin Hamiltonian takes the form
\begin{equation}\label{eq:IsingModel}
\begin{split}
H  & = \sum_{k=1}^{p-1}I^{\otimes (k-1)}\otimes (P_z)_k\otimes (P_z)_{k+1}\otimes I^{\otimes (p-k-1)} \\
   & \qquad +  \lambda \sum_{k=1}^p I^{\otimes (k-1)}\otimes (P_x)_k\otimes I^{\otimes (p-k)}\; ,
\end{split}
\end{equation}
where the index $k$ denotes the position in the spin chain and the real number $\lambda$ describes
the ratio of the strengths of the magnetic field and the pair interactions.
For simplicity, we will henceforth drop the tensor powers of the identity and tacitly assume appropriate embedding.
Then a Hamiltonian for an open-boundary 1D \defini{Heisenberg} ($XY$) model \cite{LSM61,AKLT87} reads
\begin{equation}\label{eq:Heisenbergmodel}
\begin{split}
H &= \sum_{k=1}^{p-1}\Bigl( J_x \cdot I\otimes (P_x)_k\otimes (P_x)_{k+1}\otimes I+
J_y \cdot I\otimes (P_y)_k\otimes (P_y)_{k+1}\otimes I\Bigr)\\
 	&\qquad +  \lambda \sum_{k=1}^p I\otimes (P_x)_k\otimes I \; .
\end{split}
\end{equation}
with real constants $J_x,J_y$ and $\lambda$.
Models with all coefficients being different are called \defini{anisotropic}.

Being a sum \myref{eq:Hamiltonian} of Kronecker products of structured $2 \times 2$ matrices
many Hamiltonians have special properties:
they can be multilevel-circulant (\cite{Davis94Circulant,Tyrtyshnikov00Circulant}) or skew-circulant,
diagonal or persymmetric (\cite{CantoniButler}),
which can be exploited to derive properties of the respective eigenvalues and eigenvectors.

\subsection{Computation of Ground States: Physical Background}\label{subsec:CompGroundStatesPhysicalBackGround}

A key to understand some of the motivating guidelines lies in the somewhat
striking fact that quantum dynamical systems typically evolve in a way that looks
non-generic from a mathematical point of view.
Yet the very structure of quantum dynamics paves the way to tailored parameterizations
based on tensor compression that are efficient in the sense of scaling only polynomially in physical system size.
Some of these motivating guidelines established in quantum physics may be sketched as follows:
Composing a quantum system from its components takes a joint Hilbert space that is the
tensor product of the individual Hilbert spaces. Likewise a linear operator on the
joint Hilbert space can be composed by taking sums (or weighted linear combinations)
of tensor factors (like in Eqn.~\ref{eq:Hamiltonian}).
Clearly, in general a linear combination of tensor products does not take the form of a tensor product
itself.
Thus a quantum state space grows {\em exponentially} with the number of constituents
in contrast to a classical configuration space just growing linearly.

However, correlating quantum interactions typically become smaller and smaller with
increasing distance between subsystems (\/`particles\/'): for instance, in Eqn.~\ref{eq:IsingModel}
only nearest-neighbor interactions had to be taken into account.
On a general scale,
this can be made precise in terms of {\em area laws}, where the correlations are
quantified by a measure termed \/`entanglement entropy\/' of ground states
\cite{Plenio05,Plenio06,Wolf07b}, see also the recent review in \cite{ECP10}.
Remarkably, this entropy of the reduced state of a subregion is not extensive:
it typically grows with the boundary region (\/`area\/') between the subregion and
its complement rather than with the volume of the subregion.
In one-dimensional systems, a rigorous area law has recently been proven for all
systems with a gap between the smallest and the second smallest eigenvalue \cite{Hastings07}.
Extending the results to two-dimensional lattice systems, however, currently requires
stronger assumptions on the eigenvalue distribution \cite{Hastings07b}.

Guided by these area laws, long-distance correlations may be neglected
in the sense that eigenvectors (ground states) of physical systems are
well approximated within truncated subsets, as has been quantitatively
established, e.g., for \mps \cite{VC06}.
Moreover \mps-approximations to ground states can provably be calculated efficiently \cite{Schuch09}.
Along similar lines, consecutive partitionings have been exploited in unitary and tensor
networks addressing ground states and dynamics of large-scale quantum systems.
Related techniques for truncating the Hilbert space to pertinent parameterized subsets
not only include Matrix Product States (\mps) \cite{Fannes92a, Fannes92b}
of Density Matrix Renormalization Groups ({\sc dmrg}) \cite{LNP528,Schollwoeck05}, but also
projected entangled pair states ({\sc peps}) \cite{VC04a,VC04b}, weighted graph states ({\sc wgs}) \cite{Anders06},
Multi-scale Entanglement Renormalization Approaches ({\sc mera}) \cite{Vidal07},
string-bond states ({\sc sbs}) \cite{Schuch08a} as well as combined methods \cite{Eisert07,Eisert08}.

To conclude, the evolution of physical systems does not exploit the generic state space
(with long-distance many-body interactions), but proceeds via well-defined subspaces of short-distance
and mainly pairwise interactions that can be parameterized by data-sparse formats which allow
for tensor-contraction schemes.

\subsection{Computation of Ground States: Numerical Aspects}\label{subsec:CompGroundStatesNumericalAspects}
The ground state energy of a physical system modeled by the Hamiltonian~$H$ corresponds
to the smallest eigenvalue of $H$, which is the minimum of the Rayleigh quotient \cite{HJ1}
\begin{equation}\label{eq:minimizeRayleighH}
\min_{\vec{x}\in \mathcal H}\frac{\vec{x}^{\herm}H\vec{x}}{\vec{x}^{\herm}\vec{x}} \; .
\end{equation}
As long as $p$, the number of particles, is not too large, any standard numerical method
for computing eigenvalues and eigenvectors can be used.

But with increasing $p$, the size of the Hamiltonians grows like $2^p$.
Thus for $p>50$ neither matrices nor even vectors of this size can be stored.
So, similar to the description of the Hamiltonian \myref{eq:Hamiltonian}
we need a sparse approximate representation of eigenvectors.
Assume that we have already chosen an appropriate subset $\mathcal U \subset \mathcal H$,
the goal is to find approximations for the eigenvector in this set.
Hence we consider the minimization of the Rayleigh quotient (\ref{eq:minimizeRayleighH}) only on the subset $\mathcal U$:
\begin{equation}\label{eq:minimizeRayleighU}
\min_{\vec x\in \mathcal U}\frac{\vec x^{\herm}H \vec x}{\vec x^{\herm} \vec x} \; .
\end{equation}
An appropriate subset of vectors should allow for easy computation of $H \vec x$ and of inner products
$\vec y^{\herm} \vec x$.
Therefore we consider vector representations with a less number of coefficients
where subsets of the indices can be grouped in partitions corresponding to the binary tensor structure
of the Hamiltonian~(\ref{eq:Hamiltonian}).
Please note that, in general, the chosen subsets do not form linear subspaces.
\medskip

\subsection{Alternating Least Squares}\label{subsec:als}
An important tool for minimizing the Rayleigh quotient for an appropriate subset $\mathcal U$
is the Alternating Least Squares approach (\ALS), see \cite{CarrollChang70ALS, Multi-way}.
Here, all subsets of the partitioning up to one are assumed to be fixed, and then the
minimization is reduced to the remaining subset.
As introductory example let us look at a set of vectors defined by
\begin{equation}\label{eq:phys1}
\vec{x}=\vec{x_1}\otimes \vec{x_2}\otimes \cdots \otimes \vec{x_p}
=( {x_{1;i_1}}\cdot x_{2;i_2}\cdots x_{p;i_p})_{i_1,\dots,i_p}
=(x_i)_{i=0,\cdots , 2^p-1}
\end{equation}
with vectors $\vec{x_i}$ of length $2$, and $i=(i_1,\dots, i_p)_2$ the binary representation of $i$ with $i_j\in \{ 0,1\}$.
Hence we envisage the vector $\vec{x}$ as a $p$-tensor.
So in our example (\ref{eq:phys1}) we assume all subsets fixed up to $\vec{x_r}$,
and then the minimization is simplified to
{\small{
\begin{equation*}
\begin{split}
\min_{\vec{x_r}}\frac{\vec x^{\herm}H \vec x}{\vec x^{\herm} \vec x} & =  \min_{\vec{x_r}}
            \frac{
                \left( \vec{x_1} \otimes \cdots \otimes \vec{x_p} \right)^{\herm} \left( \sum\limits_{k=1}^M \alpha_k Q_1^{(k)} \otimes \cdots \otimes Q_p^{(k)} \right) \left( \vec{x_1}\otimes \cdots \otimes \vec{x_p} \right)
                }{
                \left( \vec{x_1} \otimes \cdots \otimes \vec{x_p} \right)^{\herm} \left( \vec{x_1}\otimes \cdots \otimes \vec{x_p}\right)
                } \\
                & =  \min_{\vec{x_r}} \frac{
                    \sum\limits_{k=1}^M \alpha_k ( \vec{x_1}^{\herm} Q_1^{(k)} \vec{x_1} ) \cdots ( \vec{x_r}^{\herm} Q_r^{(k)} \vec{x_r}) \cdots (\vec{x_p}^{\herm} Q_p^{(k)} \vec{x_p})
                    }{
                    (\vec{x_1}^{\herm} \vec{x_1})\cdots ( \vec{x_r}^{\herm} \vec{x_r})\cdots(\vec{x_p}^{\herm} \vec{x_p})
                    }\\
                    & =  \min_{\vec{x_r}} \frac{ \vec{x_r}^{\herm} \left(  \sum\limits_{k=1}^M \alpha_k \beta_k Q_r^{(k)} \right) \vec{x_r}
                    }{
                    \vec{x_r}^{\herm} \left( \gamma I  \right) \vec{x_r}}
                    = \min_{\vec{x_r}}\frac{\vec{x_r}^{\herm}R_r \vec{x_r}}{ \vec{x_r}^{\herm} \vec{x_r}} \; ,\\
\end{split}
\end{equation*}
}}
a standard eigenvalue problem in the effective Hamiltonian
$R_r = \sum_{k=1}^M \tfrac{\alpha_k \beta_k}{\gamma} Q_r^{(k)}$.
More generally,
if we work with more complex representations such as Matrix Product States,
the minimization of the Rayleigh quotient~\myref{eq:minimizeRayleighU}
will lead to
\begin{equation}\label{eq:alsGenEV}
\min_{\vec{x_r}}\frac{\vec x^{\herm}H \vec x}{\vec x^{\herm} \vec x}
= \min_{\vec{x_r}}\frac{\vec{x_r}^{\herm}R_r \vec{x_r}}{ \vec{x_r}^{\herm} N_r \vec{x_r}}
\end{equation}
with an effective Hamiltonian related to the generalized eigenvalue problem in $R_r$ and $N_r$.
So far, Eq. \ref{eq:alsGenEV} describes the general situation,
the particular specification of the matrices $R_r$ and $N_r$ will be given
when we consider different representation schemes for the eigenvector.
Then, $\vec{x_r}$ is set to the eigenvector with smallest eigenvalue.
We can repeat this procedure step by step for all $\vec{x_j}$
to get approximations for the minimal eigenvalue of $H$.
The main costs are caused by matrix-vector products $H \vec x$ and inner products $\vec y^{\herm} \vec x$
plus the solution of the relatively small generalized
eigenvalue problem (\ref{eq:alsGenEV}).
It is therefore important to have efficient schemes for the evaluation of inner products of two vectors
out of the chosen subset $\mathcal U$.
We emphasize that this approach (and adapted modifications) allows to overcome
the curse of dimensionality as it is only polynomial in the maximum length of the small vectors $\vec{x_j}$,
in the number of such vectors (which can be upper bounded by $p$) and in the number of local terms $M$.

The ansatz (\ref{eq:alsGenEV}) may cause problems if the denominator matrix $N_r$ is singular.
In that case one would apply an orthogonal projection on the nonsingular subspace of $N_r$.

\section{Representations of States for Computing Ground States}
\label{sec:states}

Obviously, in general the above simplistic approach based on a single tensor
product cannot give good approximations to the eigenvector.
Therefore, we have to find a clever combination of such terms.
The choice naturally depends on the dimension and the neighborhood relation of the physical setting.
So first we consider the $1$D linear setting, and in a following section we look at the 2D problem.

\subsection{1D Systems: Approximation by Matrix Product States}\label{subsec:mps}
The Matrix Product State (\MPS) formalism goes back to several sources:
early ones are by Affleck, Kennedy, Lieb, and Tasaki
\cite{Affleck85, AKLT87} including their revival
by Fannes, Nachtergaele, and Werner \cite{Fannes92a,Fannes92b}, while
a more recent treatment is due to Vidal \cite{Vidal03}.
The application to the eigenvalue problem was discussed by Delgado et al.~\cite{Delgado01}.
A seemingly independent line of thought resorts to the fact that Density Matrix Renormalization
Group (DMRG) methods as developed by Wilson and White \cite{Wilson75, White92} have a natural
interpretation as optimization in the class of \MPS states,
see, e.g., Refs.~\cite{LNP528, Schollwoeck05, Schuch08b}.
As has been mentioned already, ground states of gapped $1$D Hamiltonians are faithfully represented
by \MPS \cite{VC06}, where the \MPS-approximation can be computed efficiently \cite{Schuch08b, Schuch09},
the rationale being an area law \cite{Hastings07}.

\subsubsection{Formalism and Computations}

For \MPS small $D_j \times D_{j+1}$-matrices are used for describing vectors in a compact form.
The advantage is due to the fact that $D:=\max\{D_j\}$ (the \emph{bond dimension}) has to grow only polynomially
in the number of particles in order to approximate ground states with a given precision (\cite{VC06}).

In \MPS, the vector components are given by
\begin{equation}\label{eq:mps}
\begin{split}
x_i = x_{i_1,\dots, i_p} & = \trace \left(A_1^{(i_1)}\cdot A_2^{(i_2)}\cdots A_p^{(i_p)} \right) \\
& = \sum_{m_1=1}^{D_1} \cdots \sum_{m_p=1}^{D_p} a_{1;m_1,m_2}^{(i_1)}\cdot a_{2;m_2,m_3}^{(i_2)}\cdot \dots \cdot
a_{p;m_p,m_1}^{(i_p)}\; .
\end{split}
\end{equation}
The matrix products lead to indices and summation over $m_2,...m_p$, and the trace introduces
$m_1$.
The upper (physical or given) indices $i_j$ identify which of the two possible matrices are used
at each position, and thereby they determine the vector components.
So, e.g., the last component is described by
\begin{equation*}
x_{2^p-1}=x_{1,\dots, 1}= \trace(A_1^{(1)}\cdot A_2^{(1)} \cdots A_p^{(1)})=
\sum_{m_1,\dots ,m_p} a_{1;m_1,m_2}^{(1)}\cdot a_{2;m_2,m_3}^{(1)} \cdots
a_{p;m_p,m_1}^{(1)}\; ,
\end{equation*}
where we always choose the matrix index $i_j = 1$.
The additional summation indices are called \emph{ancilla} indices.

The above trace form is related to the periodic case.
In the open boundary case, there is no connection between first and last particle,
and therefore the index $m_1$ can be neglected.

In this case we have $D_1 = D_{p+1} = 1$ and thus the matrices
at the ends are of size $1 \times D_2$ and $D_p \times 1$ respectively.
\begin{equation}
\begin{split}
x_{i_1,\dots,i_p} & = {A_1^{(i_1)}}\cdot A_2^{(i_2)}\cdot \dots \cdot A_{p-1}^{(i_{p-1})}\cdot A_p^{(i_p)} \\
& = \sum_{m_2=1}^{D_2}\cdots \sum_{m_p=1}^{D_p} a_{1;1,m_2}^{(i_1)}\cdot a_{2;m_2,m_3}^{(i_2)}\cdot \dots \cdot
a_{p-1;m_{p-1},m_p}^{(i_{p-1})}\cdot  a_{p;m_p,1}^{(i_p)}\; .
\end{split}
\end{equation}

By introducing the unit vectors $\vec{e_i} = \vec{e_{i_1, \dots, i_p}} = \vec{e_{i_1}} \otimes \cdots \otimes \vec{e_{i_p}}$
with unit vectors $\vec{e_{i_j}}$ of length $2$,
another useful representation of the \MPS vector is given by
\begin{equation}\label{eq:MPSasTensorProduct}
\begin{split}
\vec x & = \sum_{i_1,\cdots ,i_p}x_{i_1,\dots, i_p} \vec{e_{i_1,\dots, i_p}}=
\sum_{i_1,\cdots ,i_p} \trace(A_1^{(i_1)}\cdot \dots \cdot A_p^{(i_p)}) \vec{e_{i_1,\dots, i_p}} \\
& =\sum_{i_1,\cdots ,i_p}\sum_{m_1,\cdots ,m_p}a_{1;m_1,m_2}^{(i_1)}\cdot \dots \cdot
a_{p;m_p,m_1}^{(i_p)} \vec{e_{i_1,\dots, i_p}} \\
& =\sum_{m_1,\dots ,m_p}\Bigl (\sum_{i_1}a_{1;m_1,m_2}^{(i_1)} \vec{e_{i_1}}\Bigr )\otimes\cdots \otimes
\Bigl (\sum_{i_p}a_{p;m_p,m_1}^{(i_p)} \vec{e_{i_p}}\Bigr ) \\
& =\sum_{m_1,\cdots ,m_p} \vec{a_{1;m_1,m_2}} \otimes \cdots \otimes \vec{a_{p;m_p,m_1}}
\end{split}
\end{equation}
with length $2$ vectors $\vec{a_{r;m_r,m_{r+1}}}$ where the two components are pairwise entries
in the matrices $A_r^{(0)}$ and $A_r^{(1)}$ at position $(m_r,m_{r+1})$:
$$ \vec{a_{r;m_r,m_{r+1}}} := \begin{pmatrix}
a_{r;m_r,m_{r+1}}^{(0)} \\
a_{r;m_r,m_{r+1}}^{(1)}
\end{pmatrix} \; .
$$

\subsubsection*{Uniqueness of MPS and Normal Forms} \label{sec:uniqueness}
In this section, we want to summarize some known results concerning the uniqueness of MPS.
For further details, see, e.g., \cite{PerezGarcia07}.
Obviously the representation of an \MPS vector is not unique.
So, for a vector with components
\begin{equation}
x_i=x_{i_1,\dots,i_p}= \trace \left( {A_1^{(i_1)}}\cdot A_2^{(i_2)}\cdot \dots \cdot A_{p-1}^{(i_{p-1})}\cdot {A_p^{(i_p)} } \right)
\end{equation}
we can replace the matrices by
\begin{equation}
A_{j}^{(i_j)}\rightarrow M_j^{-1}A_j^{(i_j)}M_{j+1}\; ,A_1^{(i_1)}\rightarrow {A_1^{(i_1)} } M_2\; ,
A_p^{(i_p)}\rightarrow M_p^{-1} A_p^{(i_p)}
\end{equation}
with nonsingular matrices  $M_j \in \mathbb C^{D_j \times D_j}, j=2,\dots,p$.

The absence of uniqueness also causes problems in the solution
of the effective generalized eigenvalue problem,
because the matrix $N_r$ in Eqn.~(\ref{eq:alsGenEV}) might be positive semidefinite, but singular.
To avoid this problem, we switch from a given \MPS representation to a representation based on unitary matrices.
To this end, we combine the matrix pair $A_r^{(i_r)}$ for $i_r=0,1$ to a rectangular matrix and compute the SVD:
\begin{equation}
\mat{A_r^{(0)} \\ A_r^{(1)}} = U_r\cdot \mat{\Lambda_r \\ 0}\cdot V_r=
\mat{U_r^{(0)} \\ U_r^{(1)}}\cdot (\Lambda_rV_r)
\end{equation}
where the $U_r^{(i_r)}$ are the left part of $U_r$.
Now we can replace at position $r$ in the \MPS vector the matrix pair $A_r^{(i_r)}$ by the pair
$U_r^{(i_r)}$ and multiply the remaining SVD factor $\Lambda_r V_r$ from the left to the right
neighbor pair $A_{r+1}^{(i_{r+1})}$ without changing the vector:
\begin{eqnarray*}
\trace \left( A_1^{(i_1)}\cdot A_2^{(i_2)} \cdots A_r^{(i_r)}\cdot A_{r+1}^{(i_{r+1})} \cdot \dots A_{p-1}^{(i_{p-1})}\cdot {A_p^{(i_p)}} \right) \longrightarrow \\
\trace \left( {A_1^{(i_1)}}\cdot A_2^{(i_2)} \cdots U_r^{(i_r)}\cdot (\Lambda_r V_r) A_{r+1}^{(i_{r+1})} \cdots A_{p-1}^{(i_{p-1})}\cdot {A_p^{(i_p)}} \right) \; .
\end{eqnarray*}
So we can start from the left, normalizing first $A_1^{(i_1)}$, always moving the remaining
SVD part to the right neighbor, until we reach $A_p^{(i_p)}$.
During this procedure the \MPS matrix pairs $A_j^{(i_j)}$, $j=1,\dots,p-1$
are replaced by parts of unitary matrices
$U_j^{(i_j)}$, which fulfil the \defini{gauge condition}
\begin{equation}\label{eq:gaugeConditionLeft}
\sum_{i_j} U_j^{(i_j) \herm} U_j^{(i_j)} = I \; .
\end{equation}
In the case of open boundary conditions, the right (unnormalized) matrices $A_p^{(i_p)}$
are column vectors and thus $\sum_{i_p} A_p^{(i_p) \herm} A_p^{(i_p)}$ is only a scalar $\gamma$,
which corresponds to the squared norm of the \MPS vector:
{\allowdisplaybreaks
\begin{eqnarray*}
\vec x^{\herm} \vec x & = & \sum_{i_1,\cdots ,i_p} \overline{ \left( A_1^{(i_1)}\cdots A_p^{(i_p)} \right)} \cdot
\left( A_1^{(i_1)}\cdots A_p^{(i_p)} \right) \\
& = & \sum_{i_1,\cdots ,i_p} \left( A_1^{(i_1)} \cdots A_p^{(i_p)} \right)^{\herm} \left( A_1^{(i_1)}\cdots A_p^{(i_p)} \right) \\
& = & \sum_{i_p} A_p^{(i_p) \herm} \left(  \cdots \sum_{i_2} A_2^{(i_2) \herm}\left( \sum_{i_1} A_1^{(i_1) \herm }A_1^{(i_1)} \right) A_2^{(i_2)} \cdots \right) A_p^{(i_p)} \\
& \stackrel{\myref{eq:gaugeConditionLeft}}{=} & \sum_{i_p} A_p^{(i_p) \herm} A_p^{(i_p)} = \gamma \; .
\end{eqnarray*}
}
Thus, if $\vec x$ has norm one, the gauge condition \myref{eq:gaugeConditionLeft} is also fulfilled for $j=p$.

The same procedure can be applied in order to move the remaining SVD part to the left neighbor.
To this end, we compute
\begin{equation}
\mat{A_r^{(0)} \ A_r^{(1)}} = V_r\cdot \mat{\Lambda_r \ 0}\cdot U_r=
(V_r\Lambda_r) \mat{U_r^{(0)} \ U_r^{(1)}}\; .
\end{equation}
Similarly we can move from right to left and replace the matrix pairs $A_j^{(i_j)}$, $j=p,\dots,2$
by the unitaries $U_j^{(i_j)}$ until we reach $A_1^{(i_1)}$.
Now the gauge conditions take the form
\begin{equation}\label{eq:gaugeConditionRight}
\sum_{i_j} U_j^{(i_j)} U_j^{(i_j) \herm} = I \; .
\end{equation}
Analogously, for open boundary conditions the remaining left matrices $A_1^{(i_1)}$ are row vectors
and so $\sum_{i_p} A_1^{(i_1)} A_1^{(i_1) \herm}$ is simply a scalar, which is $1$ for a norm $1$ vector.

So far, for the normalization process only one matrix pair $A_j^{(i_j)}$ was involved.
Similar to the two-site DMRG approach~\cite{Schollwoeck05},
it is also possible to consider the matrices related to two neighboring sites at once \cite{SchollDMRG2011}.
To this end, we consider the two matrix pairs $A_j^{(i_j)} \in \mathbb C^{D_j \times D_{j+1}}$
and \mbox{$A_{j+1}^{(i_{j+1})} \in \mathbb C^{D_{j+1} \times D_{j+2}}$}.
The four matrix products $A_j^{(i_j)} A_{j+1}^{(i_{j+1})} \in \mathbb C^{D_j \times D_{j+2}}$.
are now re-arranged in matrix notation and an SVD is carried out:
\begin{equation*}
\begin{pmatrix}
    A_{j}^{(0)} \\
    A_{j}^{(1)} \\
  \end{pmatrix}
   \begin{pmatrix}
            A_{j+1}^{(0)} & A_{j+1}^{(1)} \\
          \end{pmatrix} =
                    \begin{pmatrix}
                      A_{j}^{(0)} A_{j+1}^{(0)} & A_{j}^{(0)} A_{j+1}^{(1)}  \\
                      A_{j}^{(1)} A_{j+1}^{(0)} & A_{j}^{(1)} A_{j+1}^{(1)}  \\
                    \end{pmatrix} = \begin{pmatrix}
                                U_j^{(0)} \\
                                U_j^{(1)} \\
                              \end{pmatrix} \Sigma_j \begin{pmatrix}
                                               V_{j+1}^{(0)} & V_{j+1}^{(1)} \\
                                             \end{pmatrix} \; .
\end{equation*}
If we sweep from left to right, we replace the matrices $A_j^{(i_j)}$ by parts of unitary matrices
$U_j^{(i_j)}$, shift the remaining part to the right neighbor, i.e.
$$A_{j+1}^{(i_{j+1})} \leftarrow \Sigma_j V_{j+1}^{(i_{j+1})} \; $$
and proceed with the adjacent sites $j+1$ and $j+2$.
Accordingly, if we sweep from right to left, we replace the matrices $A_{j+1}^{(i_{j+1})}$ by the unitaries
$V_{j+1}^{(i_{j+1})}$, shift the remaining part to site $j$, i.e.
$$A_{j}^{(i_{j})} \leftarrow U_{j}^{(i_{j})} \Sigma_j \;$$
and proceed with the index pair $(j-1,j)$.

There exist even stronger normalization conditions allowing representations which are
unique up to permutations and degeneracies in the singular values,
see, e.g. \cite{Eckholt11Matrix,PerezGarcia07}.
The proof of the existence of such normal forms is based on the SVD
of special matricizations of the vector to be represented, see \cite{Eckholt11Matrix, Huckle11Exploiting}.
In the SVD-TT algorithm \cite{Oseledets11tt}, the same technique is in use.
In \cite{Huckle11Exploiting} we present normal forms for \MPS which allow for expressing certain symmetry relations.

The presented normalization techniques for \MPS vectors have various advantages.
They introduce normal forms for \MPS vectors which lead to better convergence properties.
For the minimization of the Rayleigh quotient~\myref{eq:alsGenEV},
the gauge conditions circumvent the problem of bad conditioned $N_r$ matrices in the denominator
and therefore approve numerical stability.
So far, the presented normalization technique only changes the representation of the vector
but does not change the overall vector.
However, the SVD can also be used as a truncation technique.
This could be interesting if we want to keep the matrix dimensions limited by some $D=D_{\max}$.
As an example we mention the PEPS format \cite{Verstraete04a},
where such an SVD-based reduction appears, compare Subsection \ref{subsec:peps}.

\subsubsection*{Sum of \MPS Vectors}
Unfortunately, the \MPS formalism does not define a linear subspace.
According to \cite{SchollDMRG2011} the sum of two \MPS vectors
$\vec{x}$ and $\vec{y}$, which are both in PBC form, can be formulated as
{\small{
\begin{equation}\label{eq:MPSSumLargerD}
\begin{split}
\vec{x} + \vec{y} & = \sum\limits_{i_1,\dots,i_p} \mathrm{tr} \left( A_1^{(i_1)} \cdots A_p^{(i_p)} \right)
\vec{e_{i_1, \dots, i_p}}
            + \sum\limits_{i_1,\dots,i_p} \mathrm{tr} \left( B_1^{(i_1)} \cdots B_p^{(i_p)} \right) \vec{e_{i_1, \dots, i_p}} \\
        & = \sum\limits_{i_1,\dots,i_p} \mathrm{tr}
            \left[
                    \left(
                      \begin{array}{cc}
                        A_1^{(i_1)} \cdots A_p^{(i_p)} &  \\
                         & B_1^{(i_1)} \cdots B_p^{(i_p)} \\
                      \end{array}
                    \right)
            \right] \vec{e_{i_1, \dots, i_p}} \\
        & = \sum\limits_{i_1,\dots,i_p} \mathrm{tr}
            \left[
                    \left(
                      \begin{array}{cc}
                        A_1^{(i_1)} &  \\
                         & B_1^{(i_1)} \\
                      \end{array}
                    \right) \cdots \left(
                                     \begin{array}{cc}
                                       A_p^{(i_p)} &  \\
                                        & B_p^{(i_p)} \\
                                     \end{array}
                                   \right)
            \right] \vec{e_{i_1, \dots, i_p}}\\
        & = \sum\limits_{i_1,\dots,i_p} \mathrm{tr}
            \left(
                    C_1^{(i_1)} \cdots C_p^{(i_p)}
            \right) \vec{e_{i_1, \dots, i_p}} \; . \\
\end{split}
\end{equation} }}
In the open boundary case, we can define the interior matrices $C_2,\dots,C_{p-1}$ in the same way.
For reasons of consistency, the $C$ matrices at the boundary sites $1$ and $p$ have to be specified to be vectors, i.e.
\begin{equation*}
C_1^{(i_1)} = \left( A_1^{(i_1)}, B_1^{(i_1)} \right) \; ,
    \qquad C_p^{(i_p)} = \left(
                           \begin{array}{c}
                             A_p^{(i_p)} \\
                             B_p^{(i_p)} \\
                           \end{array}
                         \right) \; .
\end{equation*}
Hence, the sum of \MPS vectors is again an \MPS vector, but of larger size.
In order to keep the sizes of the \MPS matrices limited by a constant $D_{\max}$,
one could apply an SVD-based truncation of the resulting \MPS formalism and consider only the $D_{\max}$
dominant terms.

\subsubsection*{Properties of \MPS}
Every unit vector can be represented by an \MPS with bond dimension $D=1$:
let $j=(j_1,\dots,j_p)$ be the binary form of $j$, then $\vec{e_j}$
can be formulated as an \MPS with $1 \times 1$ \MPS matrices $A_r^{(i_r)} = \delta_{i_r,j_r}$:
$$ \vec{e_j} =
\sum\limits_{i_1,\dots,i_p=0}^1 \left( \delta_{i_1,j_1} \cdots \delta_{i_p,j_p} \right) \vec{e_{i_1, \dots, i_p}} \; .$$
In view of Eqn. \myref{eq:MPSSumLargerD} this observation may be extended as follows:
every sparse vector with at most $D$ non zero entries can be
written as an \MPS with bond dimension $D$.

\subsubsection*{Computation of Matrix-Vector Products}
To solve the minimization of the Rayleigh quotient (\ref{eq:alsGenEV}) in an \ALS way
as introduced in Subsection \ref{subsec:als}, we have to compute
\begin{equation*}
\begin{split}
\vec{y_k}=H^{(k)} \vec x & =\Bigl (\alpha_k Q_{1}^{(k)} \otimes\cdots\otimes Q_{p}^{(k)} \Bigr )\cdot
 \sum_{m_1,\dots ,m_p} \Bigl ( \vec{a_{1;m_1,m_2}}\otimes\cdots\otimes \vec{a_{p;m_p,m_1}}\Bigr ) \\
& = \sum_{m_1,\dots ,m_p} \alpha_k \Bigl (Q_{1}^{(k)} \vec{a_{1;m_1,m_2}}\Bigr )\otimes\cdots\otimes
\Bigl (Q_{p}^{(k)} \vec{a_{p;m_p,m_1}}\Bigr ) \\
& = \sum_{m_1,\dots,m_p} \alpha_k \Bigl (\vec{b_{1,k;m_1,m_2}}\otimes\cdots\otimes \vec{b_{p,k;m_p,m_1}}\Bigr )
\end{split}
\end{equation*}
and derive a sum of \MPS vectors
\begin{equation}\label{eq:HamiltonianxMPSeqSumMPS}
\vec y=H \vec x=\sum_{k=1}^M H^{(k)} \vec x= \sum_{k=1}^M \vec{y_k}\; .
\end{equation}
Therefore, we can compute all small vectors $Q_{j}^{(k)} \vec{a_{j;m_j,m_{j+1}}}$ in $\mathcal O(2pD^2M)$ operations.
For reasons of efficiency there are also concepts
to express the Hamiltonian in terms of a \defini{Matrix Product Operator} (\MPO) as defined in (\ref{eq:mpdo}).
This approach enables short representations for \MPO $\times$ \MPS products.

\subsubsection*{Computation of Inner Products}
\begin{figure}[ht]
\centering
\subfigure[Computation of the inner product (\ref{eq:innerProductMPS}) of two \MPS vectors.]{
\includegraphics[width=0.45\textwidth]{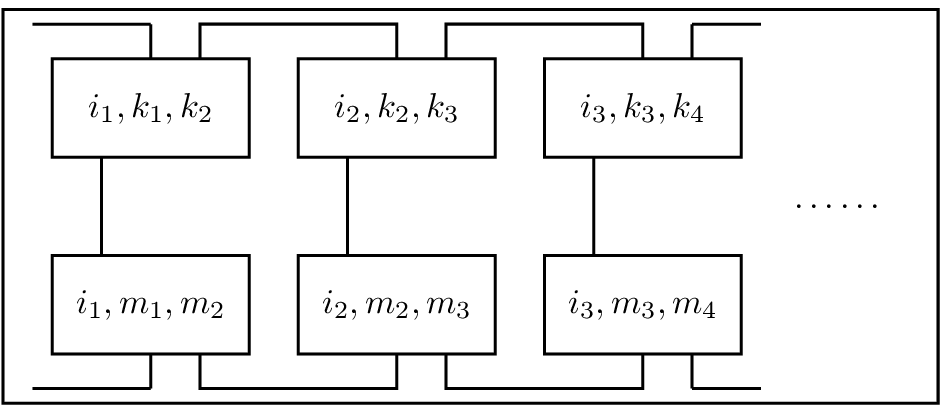}
\label{fig:mps1}
}
\hfill
\subfigure[Contraction of the two \MPS vectors concerning index $i_1$.]{
\includegraphics[width=0.45\textwidth]{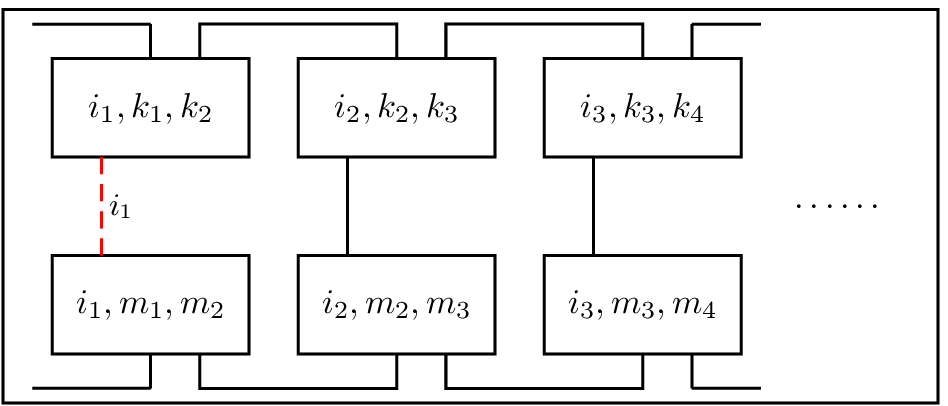}
\label{fig:mps2}
}
\caption{Contraction of the inner product of two \MPS vectors.
The dashed red line illustrates the index being contracted.}
\label{fig:contractMPS}
\end{figure}

Furthermore, we have to compute inner products of two \MPS vectors
\begin{equation}\label{eq:innerProductMPS}
\sum_{i_1,\dots,i_p}\sum_{\begin{subarray}{c} m_1,\dots , m_p \\ k_1, \dots
,k_p \end{subarray}}\Bigl (\bar b_{1;k_1,k_2}^{(i_1)}\cdot \dots \cdot
\bar b_{p;k_p,k_1}^{(i_p)}\Bigr )\cdot \Bigl (a_{1;m_1,m_2}^{(i_1)}\cdot \dots \cdot
a_{p;m_p,m_1}^{(i_p)}\Bigr )\; .
\end{equation}
To deal with these sums, it is essential to specify in which order the summations have to be conducted.
In the next figures we display such an efficient ordering for the periodic case.
To this end, we introduce the following notation.
Each box in the following figures describes one factor, e.g., $a_{r;m_r,m_{r+1}}^{(i_r)}$, in these
collections of sums.
Little lines (legs) describe the connection of two such boxes via a common index.
Hence, the number of legs of a box is exactly the number of indices.
So in this case, most of the boxes have three legs.
Figure \ref{fig:mps1} shows the sum and the first terms, resp. boxes, with their connections.
This also represents a Tensor Network.

\begin{figure}[ht]
\centering
\subfigure[After the $i_1$-contraction, we get a four leg tensor]{
\includegraphics[width=0.45\textwidth]{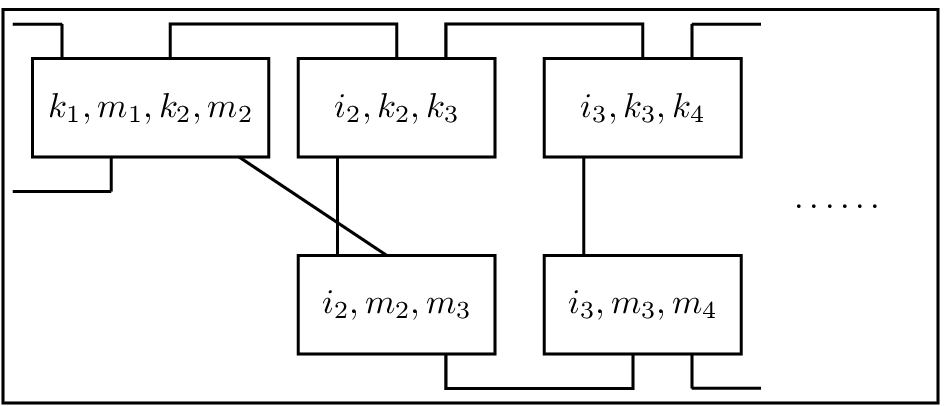}
\label{fig:mps3}
}
\hfill
\subfigure[Contraction concerning index $k_2$]{
\includegraphics[width=0.45\textwidth]{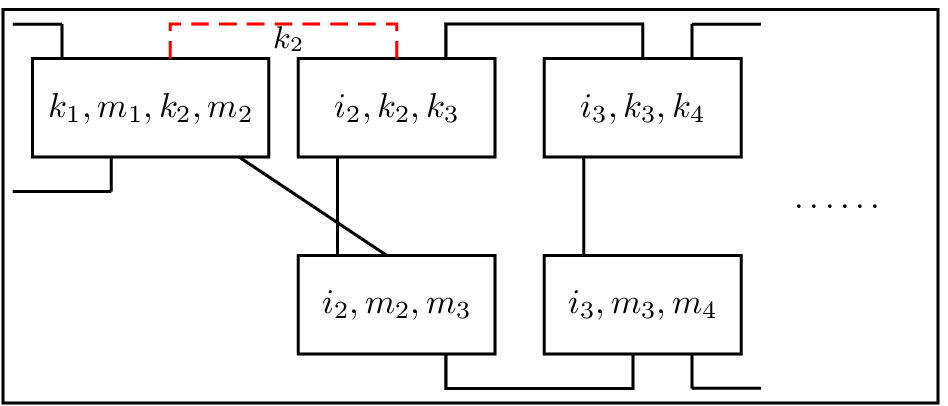}
\label{fig:mps4}
}
\vfill
\subfigure[Contraction concerning the indices $i_2$ and $m_2$]{
\includegraphics[width=0.45\textwidth]{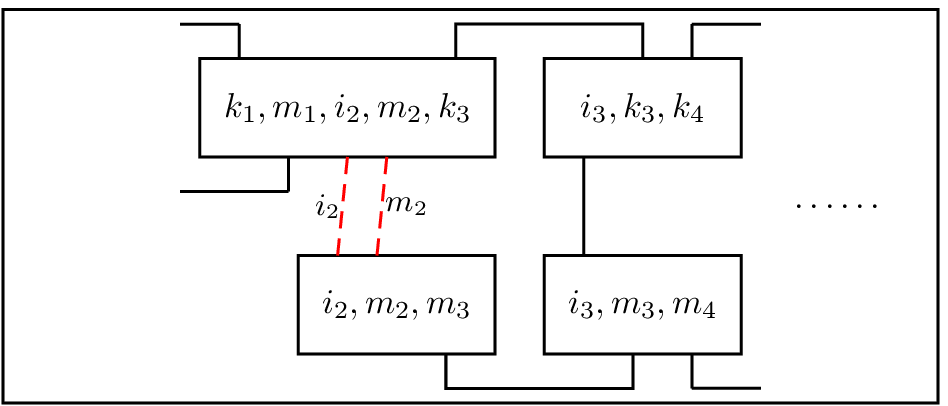}
\label{fig:mps5}
}
\hfill
\subfigure[After the contraction concerning $k_2, m_2$ and $i_2$, we are in the same situation
as in Figure \ref{fig:mps3}]{
\includegraphics[width=0.45\textwidth]{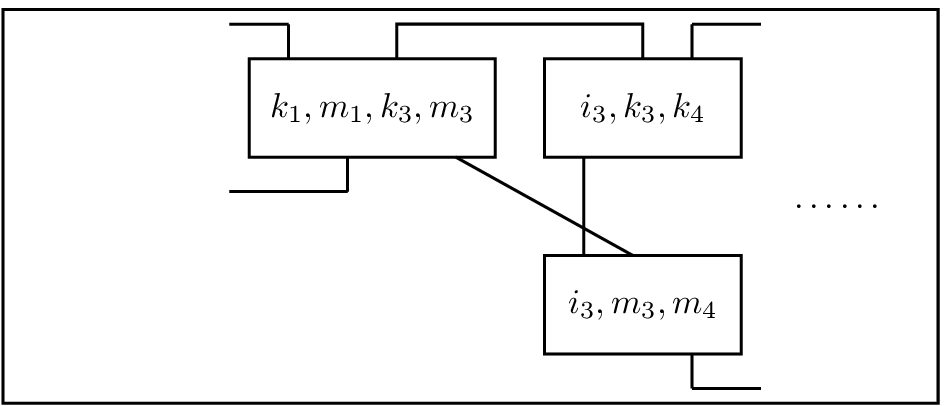}
\label{fig:mps6}
}
\caption{Contraction of the inner product of two \MPS vectors.}
\label{fig:contractMPS2}
\end{figure}

Now, in a first step we reorder the sums, and execute a contraction relative to index $i_1$.
This is given by the partial sum
\begin{equation}
\sum_{i_1} \bar b_{1;k_1, k_2}^{(i_1)} a_{1;m_1, m_2}^{(i_1)}=c_{k_1,k_2;m_1,m_2} \ .
\end{equation}
This eliminates two boxes, but leads to a new four leg tensor $c_{k_1,k_2;m_1,m_2}$ as shown
in \ref{fig:mps2} and \ref{fig:mps3}.
Now we contract index $k_2$, as shown in Figure \ref{fig:mps4}, leading to \ref{fig:mps5}.
In the following step we go from \ref{fig:mps5} to \ref{fig:mps6} by contracting $i_2$ and
$m_2$, deriving at Figure \ref{fig:mps6}.
Now we are in the same situation as in Figure \ref{fig:mps3}, and we can proceed exactly
in the same way, until all sums are executed, resp. all indices have been contracted.

The main costs depend on the size of the contracted index, e.g. $2$ for contracting $i_r$ or
$D$ for contracting $m_r$ or $k_r$, and on the size of the other indices that appear in
the contracted boxes.
Hence, e.g. the contraction in Figure \ref{fig:mps4} costs $D\cdot 2D^3=2D^4$ operations,
and Figure \ref{fig:mps5} costs $2D\cdot D^4=2D^5$ operations.
The total costs for the inner product is therefore less than $4D^5p$, because contractions
have to be done until all boxes are removed, that is $2p$-times.
Therefore, the costs for computing $\vec x^{\herm}H \vec x$ based on \myref{eq:HamiltonianxMPSeqSumMPS}
are less than $4D^5Mp$.

In the open boundary case, we start at the left or right side and therefore only contractions of boxes
of length $3$ can occur and thus Figure \ref{fig:mps5} only costs $2 D \cdot D^2 = 2 D^3$
instead of $2 D^5$ in the periodic case.
Thus, the overall costs for the inner product is less than $4 D^3 p$
and for $\vec x^{\herm} H \vec x$ less than $4 D^3M p$.

\subsubsection*{Minimization of the Rayleigh Quotient in Terms of \MPS}
Now, we use the \MPS formalism as vector ansatz to minimize the Rayleigh quotient \myref{eq:minimizeRayleighU}.
Therefore we start the \ALS procedure, updating $A_p^{(i_p)}$ by an improved estimate by solving
the generalized effective eigenvalue problem.
In addition, we replace $A_p^{(i_p)}$ by a unitary pair via SVD, moving the remaining SVD
term to the left neighbor, and so on.
A nice side effect here is that in the case of open boundary conditions
the matrix $N_r$ is the identity because all the matrix pairs
to index different from $r$ are parts of unitary matrices and thus fulfil one of the gauge conditions \myref{eq:gaugeConditionLeft} or \myref{eq:gaugeConditionRight}.
Together with the fact that the \MPS matrices at both ends are simply vectors we obtain
%
{\allowdisplaybreaks
\begin{eqnarray*}
\vec x^{\herm} \vec x & = & \sum_{i_1,\cdots ,i_p} \trace(\bar A_1^{(i_1)}\cdots \bar A_p^{(i_p)})\cdot
\trace(A_1^{(i_1)}\cdots A_p^{(i_p)}) \\
& = & \sum_{i_1,\cdots ,i_p} \trace\Bigl ((\bar A_1^{(i_1)}\cdots \bar A_p^{(i_p)})\otimes
(A_1^{(i_1)}\cdots A_p^{(i_p)})\Bigr ) \\
& = & \sum_{i_1,\cdots ,i_p} \trace\Bigl ((\bar A_1^{(i_1)}\otimes A_1^{(i_1)})\cdots
(\bar A_p^{(i_p)}\otimes A_p^{(i_p)})\Bigr ) \\
& = & \trace\Bigl (\sum_{i_1,\cdots ,i_p} \Bigl ((\bar A_1^{(i_1)}\otimes A_1^{(i_1)})\cdots
(\bar A_p^{(i_p)}\otimes A_p^{(i_p)})\Bigr )\Bigr ) \\
& = & \trace \Bigl ( \prod_j\sum_{i_j}(\bar A_j^{(i_j)}\otimes A_j^{(i_j)} )\Bigr ) \\
& = & \sum_{i_r} \trace\left( A_r^{(i_r)} A_r^{(i_r) \herm} \right) \; .
\end{eqnarray*}
}
In the case of periodic boundary conditions the SVD-based normalization can only be performed
for all up to one matrix pair $A_r^{(i_r)}$ and so the denominator matrix $N_r$ is non-trivial.
However, the normalization is utilized to make the problem numerically stable (see \cite{Verstraete04a}).

The matrix $H_r$ for the eigenvalue problem is given by
{\allowdisplaybreaks
\begin{align*}
\lefteqn{\vec x^{\herm} H \vec x  = }\\
& = \sum_{ \begin{subarray}{c}i_1,\dots ,i_p \\ i_1',\dots ,i_p' \\ k \end{subarray} }
\trace(\bar A_1^{(i_1)}\cdots \bar A_p^{(i_p)})\cdot \trace(A_1^{(i_1')}\cdots A_p^{(i_p')})\cdot
(\vec{e_{i_1'}}^{\herm} Q_{1}^{(k)} \vec{e_{i_1}})\cdots (\vec{e_{i_p'}}^{\herm} Q_{p}^{(k)} \vec{e_{i_p}}) \\
& =  \sum_{\begin{subarray}{c}i_1,\dots ,i_p \\ i_1',\dots ,i_p' \\ k \end{subarray}}
\trace\Bigl ((\bar A_1^{(i_1)}\cdots \bar A_p^{(i_p)})\otimes
(A_1^{(i_1')}\cdots A_p^{(i_p')})\Bigr )\cdot
\Bigl (Q_{1;i_1',i_1}^{(k)} \cdots Q_{p;i_p',i_p}^{(k)} \Bigr ) \\
& =  \sum_{\begin{subarray}{c}i_1,\dots ,i_p \\ i_1',\dots, i_p' \\ k \end{subarray}}
\trace\Bigl ((\bar A_1^{(i_1)}\otimes A_1^{(i_1')})\cdots
(\bar A_p^{(i_p)}\otimes A_p^{(i_p')})\Bigr )\cdot
\Bigl (Q_{1;i_1',i_1}^{(k)} \cdots Q_{p;i_p',i_p}^{(k)}\Bigr ) \\
& =  \sum\limits_k \trace \bigg[ \sum_{\begin{subarray}{c}i_1,\dots ,i_p \\ i_1',\dots, i_p' \end{subarray}}
     \Bigl (Q_{1;i_1',i_1}^{(k)} \cdot (\bar A_1^{(i_1)}\otimes A_1^{(i_1')}) \Bigr) \cdots
\Bigl (Q_{p;i_p',i_p}^{(k)} \cdot (\bar A_p^{(i_p)}\otimes A_p^{(i_p')})\Bigr ) \bigg ] \\
& =  \sum\limits_k \trace \Bigl ( \prod_j\sum_{i_j',i_j}(Q_{j;i_j',i_j}^{(k)} \cdot
     (\bar A_j^{(i_j)}\otimes A_j^{(i_j')} ) ) \Bigr ) \; .
\end{align*}
}
The effective Hamiltonian can be computed by contracting all indices except the indices
representing the unknowns in matrix pair $A_r^{(i_r)}$,  $i_r,i_r',m_r,m_{r+1},k_r,k_{r+1}$
leading to a $2D^2 \times 2D^2$ matrix $H_r$.
So the costs for solving the eigenvalue problem are in this case $O(D^4)$.
In the periodic case also a SVD for $N_r$ has to be computed to solve the generalized eigenvalue problem
numerically stable, which leads to costs of order $D^6$.

\subsubsection{Matrix Product Density Operators}
The Matrix Product Operator approach extends the idea behind \MPS from vectors to operators.
Matrix Product Density Operators \MPDO \cite{MPDO2004} have the form
\begin{equation}\label{eq:mpdo}
\sum_{i,i'} \trace \left( A_1^{(i_1,i'_1)}\cdot \dots \cdot A_p^{(i_p,i'_p)} \right) \vec{e_i} \vec{e_{i'}}^T
\end{equation}
with unit vectors $\vec{e_i}$.
Matrix Product Operators \MPO \cite{PMCV10} read
\begin{equation}\label{eq:mpo}
\sum\limits_{i_1,\dots,i_p} \trace \left( A_1^{(i_1)}\cdot \dots \cdot A_p^{(i_p)} \right) P_{i_1}\otimes \cdots \otimes P_{i_p}
\end{equation}
with $2 \times 2$ matrices $P$, e.g. the Pauli matrices (\ref{eq:Pauli}).
Similarly, the \TT format has also been extended to the matrix case (\cite{Oseledets10ttm}).
These \MPO concepts may be used for a representation of the Hamiltonian,
such that the application of the Hamiltonian on an \MPS vector leads to a sum of \MPS vectors with less addends.

\subsection{2D Systems: Approximation by Projected Entangled Pair States}\label{subsec:peps}
It is natural to extend the coupling topology of interest
from linear chains to 2D arrays representing an entire lattice
with open or periodic boundary conditions.
To this end, the matrices in \MPS are replaced by higher-order tensors thus
giving rise to Projected Entangled Pair States (\PEPS) \cite{VC04a}.
Again, the underlying guideline is an area law \cite{VWPGC06,Hastings07b}.
The extension to higher dimensions, however, comes at considerably higher
computational cost: calculating expectation values becomes NP-hard
(actually the complexity class is $\#P$-complete) \cite{Schuch07}.
This is one reason, why computing ground states in two-dimensional
arrays remains a major challenge to numerics \cite{VCM09}.

\begin{figure}[ht]
\begin{center}
\includegraphics[width=0.55\textwidth]{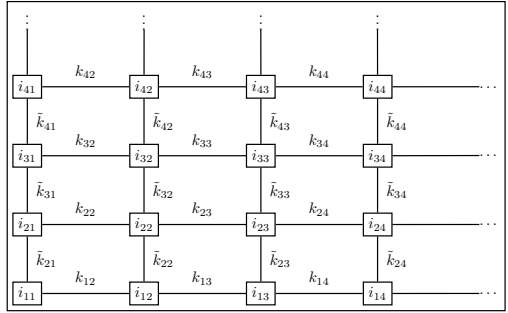}
\end{center}
\caption{A 2D system in the open boundary case with physical indices $i_{j}$ and ancilla
indices $k$ and $\tilde k$.}
\label{fig:peps1}
\end{figure}

For 2D spin systems, the interaction between particles is also of 2D form, e.g., as described
by Figure \ref{fig:peps1}.
This leads to 2D generalization of \MPS using a tensor network with small boxes related to
tensor $\left( a_{k_r,k_{r+1}; \tilde k_s, \tilde k_{s+1}}^{(i_{r,s})} \right)$ with $5$ legs.
Thus, an inner product of two such vectors would look like
\begin{equation*}
\sum \bar b_{k_r',k_{r+1}'; \tilde k_s',\tilde k_{s+1}'}^{(i_{r,s})}
          a_{k_r,k_{r+1}; \tilde k_s, \tilde k_{s+1}}^{(i_{r,s})} \; .
\end{equation*}
In a \PEPS vector, matrices are replaced by higher order tensors with one physical index related
to the given vector $\vec x$ and the other ancilla indices related to nearest neighbors according to
the lattice structure of the tensor network (\cite{PEPS2007}).
A \PEPS vector can be formally written as
\begin{equation*}
\vec x=\sum_{I_R}C_R(\{ A^{I_R}\} ) \vec{e_{I_R}}\; ,
\end{equation*}
where $I$ stands for all $\vec x$-indices in Region $R$, $C_R$ represents the contraction of indices
following the nearest neighbor structure.

\begin{figure}[htp]
\centering
\subfigure[Contracting physical indices.]{
\includegraphics[width=0.47\textwidth]{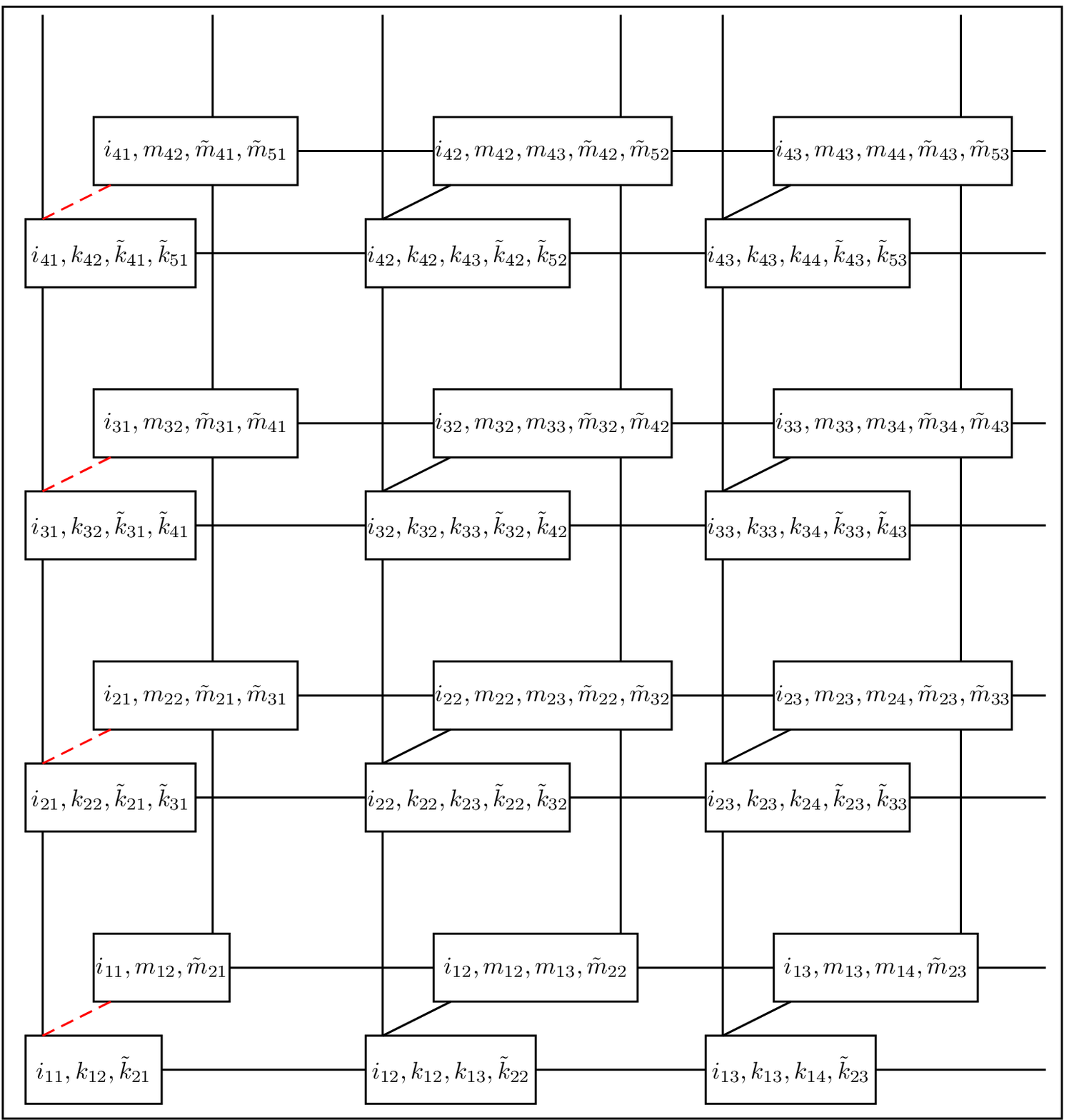}
\label{fig:peps2}
}
\subfigure[Grouping index pairs.]{
\includegraphics[width=0.47\textwidth]{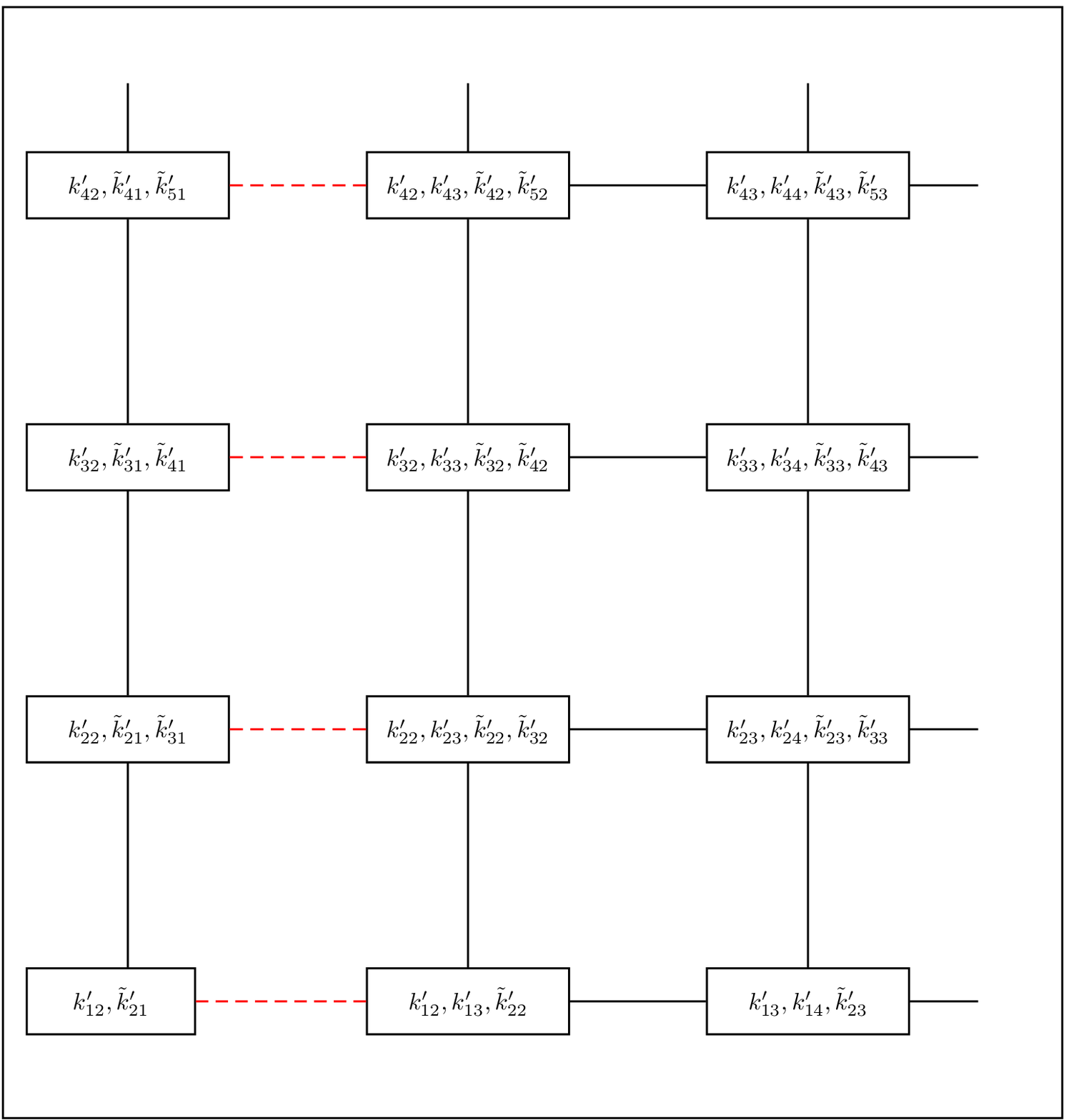}
\label{fig:peps3}
}
\caption{Contraction of two \PEPS vectors:
After the contraction of the physical indices related to the first column (a)
the index pairs $m_{r,s}$ and $k_{r,s}$ are grouped to larger indices $k_{r,s}'$.}
\label{fig:pepsContraction1}
\end{figure}

To compute the inner product of two \PEPS vectors, we have to find an efficient ordering of the summations.
The related tensor network for the open boundary case is displayed in Figure \ref{fig:peps2}.

In a first step all pairwise contractions relative to the vector indices $i_{r,s}$ are computed.
Furthermore, in the produced boxes the indices are newly numbered by combining $k_{r,s}$ with
$m_{r,s}$ to larger index $k'_{r,s}$.
This generates Figure \ref{fig:peps3}.

\begin{figure}[htp]
\subfigure[Contracting index pairs.]{
\includegraphics[width=0.48\textwidth]{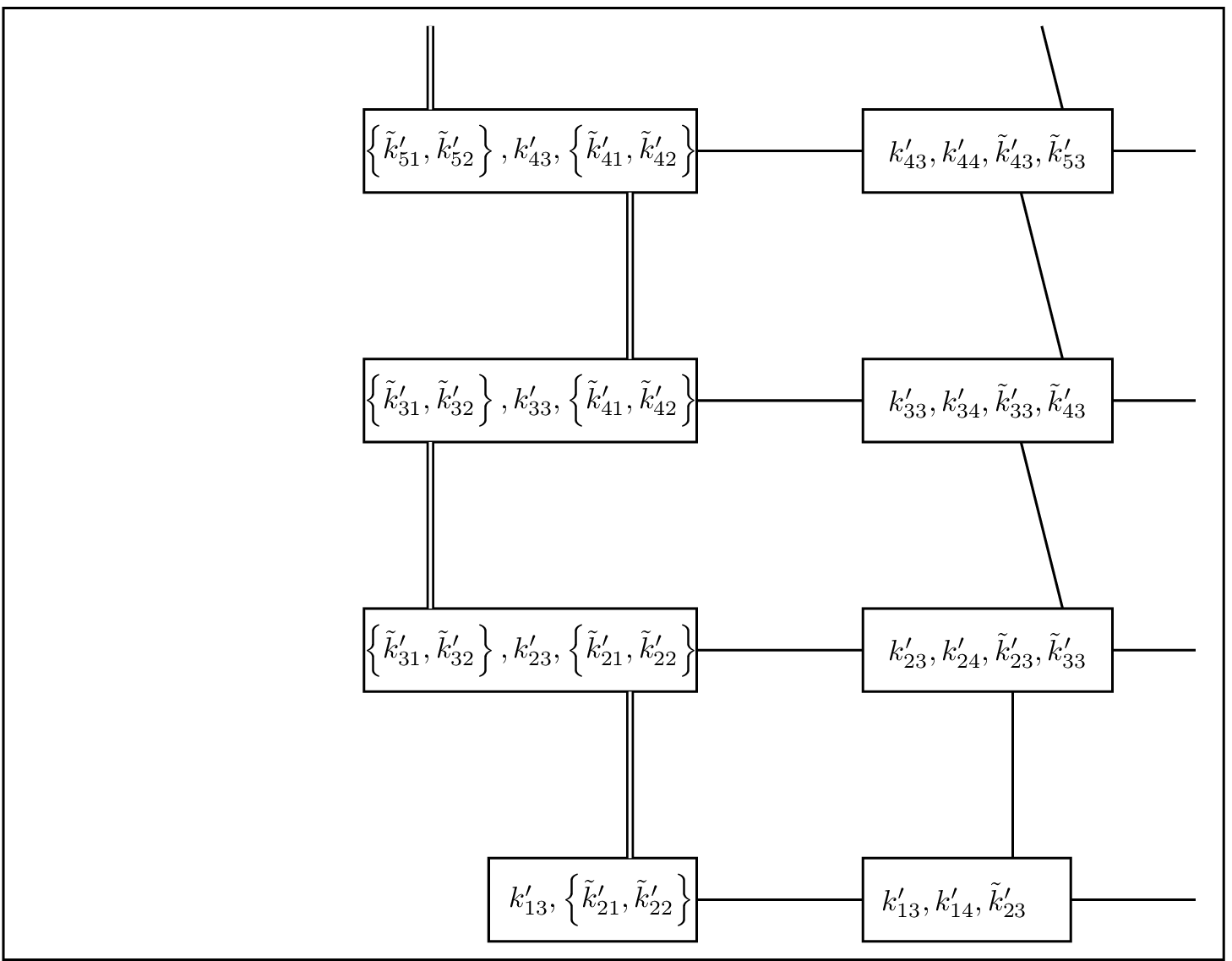}
\label{fig:peps4}
}
\hfill
\subfigure[Result of the rank reduction.]{
\includegraphics[width=0.48\textwidth]{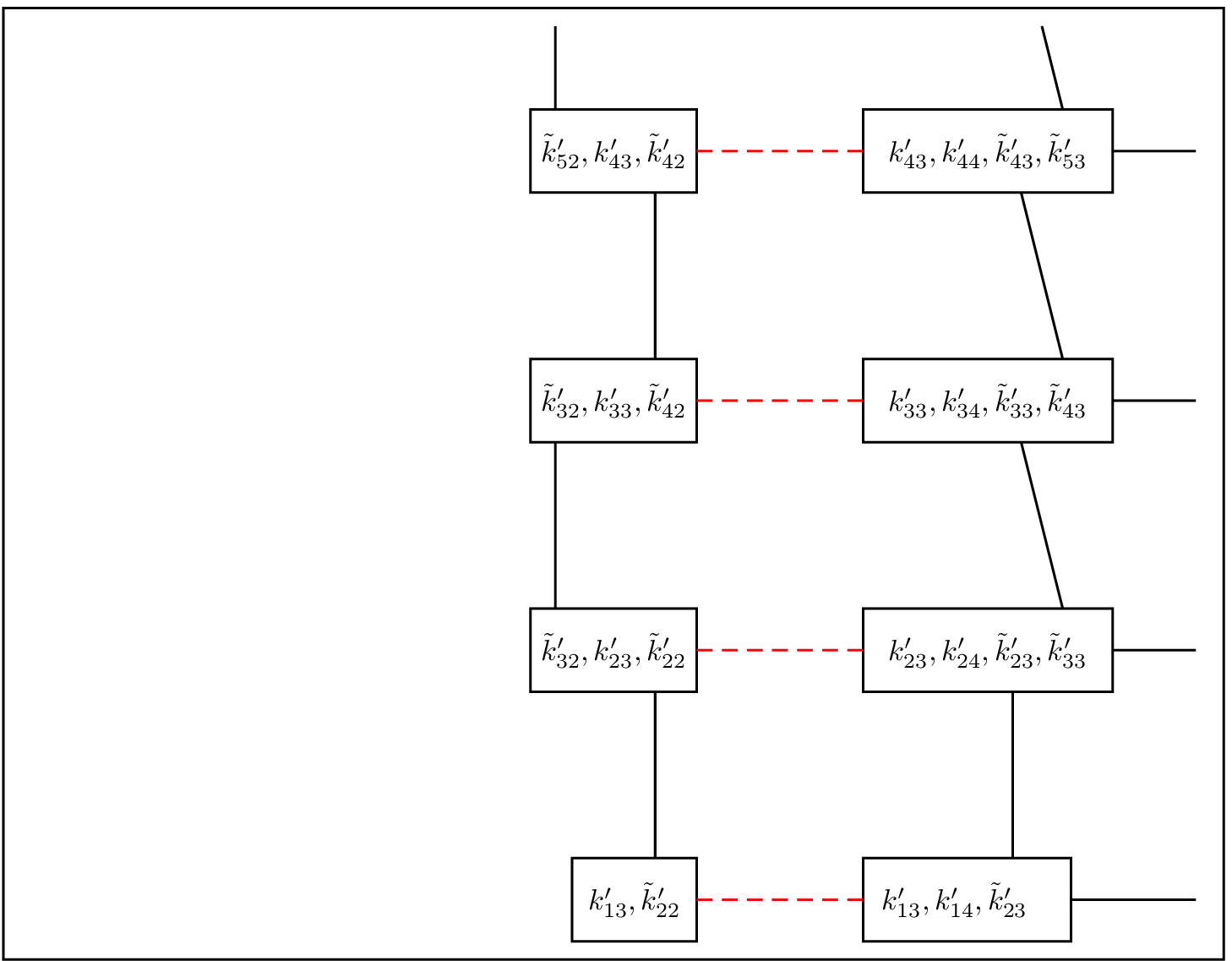}
\label{fig:peps5}
}
\caption{Contraction scheme for \PEPS vectors: after the contracting illustrated in Fig. \ref{fig:peps3},
The newly generated first column in (a) has sets of indices
leading to a more complex contraction process as illustrated by the double lines.
After the rank reduction (b), we are in the same situation as in Fig. \ref{fig:peps3} and
can thus proceed in the same manner.}
\label{fig:pepsContraction2}
\end{figure}

Now the first and second column are contracted starting e.g. from the bottom, resulting in the
network displayed in Figure \ref{fig:peps4}.
Unfortunately, the boxes in the newly generated left column have more indices than in the starting column.
So we cannot proceed in this way directly.
In order to keep the number of indices constant, an approximation step is introduced.
The main idea is to reduce the number of indices by considering the first
left column as \MPS---with indices related to connections with the right neighbors as original physical
indices (longer than $2$)---and approximating the boxes by little tensors with only one leg instead
of two to the vertical neighbors.

Such a reduction can be derived by the following procedure.
We want to reduce the rank of size $D^2$ in index pair $\{ k_{2,1}, k_{2,2}\} $ to a new index $k'_{2,1}$
of size $D$.
We can rewrite the whole summation in three parts, where $c$ contains the contracted summation
over all indices that are not involved in the actual reduction process.
This leads to the sum
\begin{equation*}
\sum a_{\{ k'_{2,3},k'_{2,1}\} ,\{ k'_{3,1},k'_{3,2}\} }\cdot
     b_{ \{k'_{3,1},k'_{3,2}\} ,\{ k'_{3,3},k'_{4,1},k'_{4,2} \} }\cdot
     c_{\{ k'_{3,3},k'_{4,1},k'_{42}\} , \{k'_{2,3},k'_{2,1}\} } \; .
     \end{equation*}
The entries $a$ in the above sum build the \MPS matrices
$A_{({k'_{2,1},k'_{2,2}}),(k'_{3,1},k'_{3,2})}^{(k_{2,3}')}$.
Collecting these matrices in a large rectangular matrix, we can apply the SVD on this matrix.
Now we can truncate the diagonal part of the SVD to reduce the matrices $A$ from size $D^2\times D^2$
to size $D^2\times D$.
If we repeat this process along the first column all indices are reduced to length $D$.
Note that this approximation step is not affected by any accuracy demand
but it is performed to keep the number of indices constant,
which allows to proceed in an iterative way.

Reducing the rank of the indices in the first column leads to the network illustrated by Figure \ref{fig:peps5}
with the same structure as Figure \ref{fig:peps3}, so we can contract column by column until
we are left with only one column that can be contracted following Figure \ref{fig:peps6}.
\begin{figure}[ht]
\begin{center}
\includegraphics[width=0.55\textwidth]{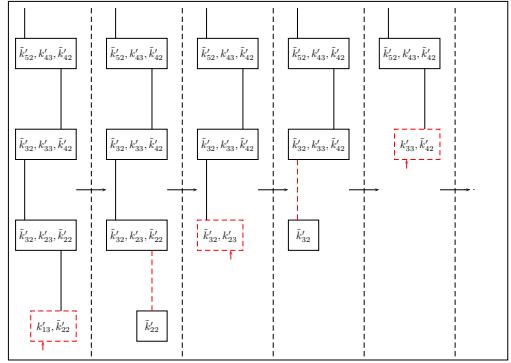}
\end{center}
\caption{After application of the column-by-column contraction scheme,
we end up with the last column, which can be contracted in the illustrated way.}
\label{fig:peps6}
\end{figure}

The overall costs depend on $D^{10}$, resp. $D^{20}$ in the periodic case.
The so-called virtual dimension $D$ is usually chosen a priori by the physicists (e.g. $D\leq 5$),
such that the computations can still be afforded (\cite{Verstraete04a}).

\subsection{Tree Tensor States and Multi-scale Entanglement Renormalization Ansatz}\label{subsec:mera}
\begin{figure}
\begin{center}
\includegraphics[width=0.75\textwidth]{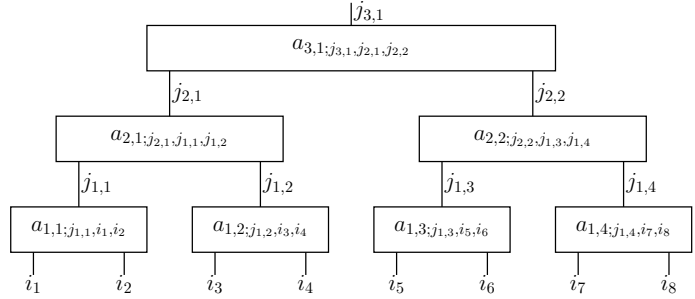}
\end{center}
\caption{Scheme of the Tree Tensor States ({\sc{tts}}).}
\label{fig:tts}
\end{figure}

In \MPS the new spin particles are added one by one leading to a new matrix in the tensor ansatz.
A modified procedure leads to \defini{Tree Tensor States} ({\sc{tts}}).
In this ansatz the particles are grouped pairwise leading to a smaller subspace
of the original Hilbert space.
This procedure is repeated with the blockpairs until only one block is left.
This is displayed in Figure \ref{fig:tts} and formula
\begin{align*}
& x_{i_1,\dots,i_8} = \\
& \sum_{j_{1,1},\dots ,j_{2,2}}{
    ( a_{1,1;j_{1,1},i_1,i_2}\cdots a_{1,4;j_{1,4},i_7,i_8})
    (  a_{2,1;j_{2,1},j_{1,1},j_{1,2}} a_{2,2;j_{2,2},j_{1,3},j_{1,4}})
   ( a_{3,1;j_{3,1},j_{2,1},j_{2,2}} )}
\end{align*}
where each block is isometric:
$$\sum_{i,j}\bar a_{k',i,j} \cdot a_{k,i,j} = \delta_{k',k}\; .$$
Hence, in this scheme the network consists in a binary tree built by small isometric tensors with three indices.
It can be shown that all \MPS can be represented by {\sc{tts}} (see \cite{VC10}).

\begin{figure}[ht]
\begin{center}
\includegraphics[width=0.7\textwidth]{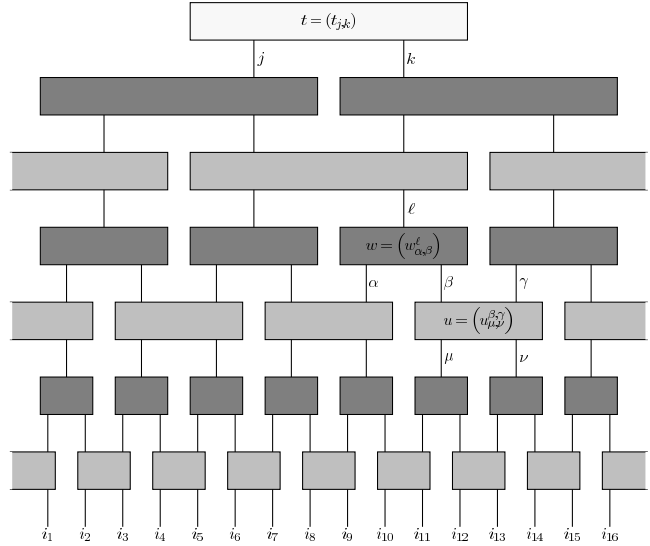}
\end{center}
\caption{The Multi-scale Entanglement Renormalization Ansatz ({\sc{mera}})
with periodic boundary conditions,
consisting of isometries $w$ and unitaries $u$.}
\label{fig:mera}
\end{figure}

A further generalization of \TTS and \MPS is given by
the \defini{Multi-Scale Entanglement Renormalization Ansatz} ({\sc mera}, see \cite{Vidal07}).
Besides the top tensor~$t$ with two indices, the network is built by two types of smaller tensors:
three leg isometric tensors (\defini{isometries}) and four leg unitary tensors (\defini{disentanglers}).
This formalism is displayed by Figure~\ref{fig:mera}.
In connection with eigenvalue approximation for {\sc{mera}}, \ALS cannot be used in order to optimize
over the isometries that represent the degrees of freedom;
instead other optimization methods have to be applied.

\section{Representations of Tensors}
\label{sec:tensor}

As any state in a physical system (see Sec. \ref{sec:physicalModelSystem}) with $p$ particles
may be seen as a $p$-th order tensor, we want to present some basics about tensors within this section.
In the next section, we will give some modifications and generalizations.

\subsection{Some Basics about Tensors}\label{subsec:BasicsTensors}
In its most general form, a $p$th-order tensor $A = ( a_{i_1, \dots, i_p}) \in \mathbb R^{n_1 \times \cdots \times n_p}$
is a multiway array with $p$ indices.
A first-order tensor is thus a vector, a second-order tensor corresponds to a matrix.
If $\vec x$ and $\vec y$ are vectors (i.e. first-order tensors) it is well known that the \defini{outer product}
$ \vec x \circ \vec y := \vec x \vec {y}^{\operatorname T}$ is a rank-one matrix.
As generalization, if $\vec{a_{1}},\dots,\vec{a_{p}}$ are vectors, the tensor
$A := \vec{a_{1}} \circ \cdots \circ \vec{a_p}$, which is defined as
$a_{i_1,\dots,i_p} = a_{1;i_1} a_{2;i_2} \cdots a_{p;i_p}$ is a \defini{rank-one tensor}.
Hence, the application of outer products constructs tensors of higher order.
The Kronecker product of matrices just corresponds to this definition of the outer product,
but it reshapes the single tensor entries into a matrix of larger size.

To begin, we write a tensor $A$ as a sum of rank-one tensors:
\begin{equation}\label{eq:sumrank1tensors}
A = \sum\limits_{j=1}^R \vec{a_1^{(j)}} \circ \vec{a_2^{(j)}} \circ \cdots \circ \vec{a_p^{(j)}} \; .
\end{equation}
If $R$ is minimal in the representation (\ref{eq:sumrank1tensors}) of $A$,
we define the \defini{tensor rank} of $A$ as equal to $R$.

For illustrating tensors, matrices are often in use.
One possibility to bring back a general tensor to a matrix is given by the \defini{mode-n-unfolding}
(see \cite{TensorReview}):
When applying this technique, the tensor $A=(a_{i_1,\dots,i_n,\dots,i_p})$ is represented by the matrix
$$A_{(n)} := \left( a_{i_1,\dots,i_n,\dots,i_p} \right)_{i_n,\{i_1,\dots,i_{n-1}, i_{n+1},\dots,i_p\}} \; .$$

The \defini{n-mode tensor matrix product} is given as follows:
Let $A = (a_{i_1,\dots,i_n,\dots,i_p})$ be a $p$-th order tensor and $U = (u_{j,i_n})$ a matrix,
then the \defini{mode-$n$-product} $\times_n$ is defined as
\begin{equation*}
A \times_n U =
\left(  \sum\limits_{i_n} a_{i_1,\dots,i_n} u_{j,i_n} \right)_{i_1,\dots,i_{n-1},j,i_{n+1},\dots,i_p} \; .
\end{equation*}


Beside the total rank of a tensor there also exist a local rank concept:
the \defini{mode-n-rank} of a tensor is the rank of the collection of all vectors belonging to index $n$.
If a given tensor $A$ has the $n$-mode ranks $r_n$, we define the \defini{multilinear rank} of $A$ as $(r_1,\dots,r_p)$.

\subsection{Decomposition of Tensors}\label{subsec:DecompTensors}
For approximating tensors $(x_{i_1,\dots,i_p})$ there are two basic methods (see \cite{TensorReview}):
the \defini{Tucker decomposition} and the \defini{canonical decomposition}.
The Tucker decomposition (\cite{bestrank-1})
\begin{equation}\label{eq:Tucker}
x_{i_1,\dots,i_p}=\sum_{m_1,\dots ,m_p}^D y_{m_1,\dots ,m_p}a_{1;i_1,m_1} a_{2;i_2,m_2} \cdot \dots \cdot a_{p;i_p,m_p}
\end{equation}
represents the given tensor by a tensor $y_{m_1,\dots ,m_p}$ with less dimension in each direction;
$D$ is called the \defini{Tucker rank}, $y_{m_1,\dots ,m_p}$ is called \defini{core tensor}.
This concept is illustrated in Figure \ref{fig:Tucker}.
In the case of a binary tensor $\vec x \in \mathbb R^{2 \times \cdots \times 2}$, this is not meaningful,
because then the Tucker rank is already $2$.

The canonical decomposition (candecomp), which is also known as Parallel Factorization(\PARAFAC),
has the form
\begin{equation}\label{eq:canonicalDecomposition}
x_{i_1,\dots,i_p}=\sum_{s=1}^D a_{1;i_1}^{(s)} a_{2;i_2}^{(s)} \cdot \dots \cdot a_{p;i_p}^{(s)} \; .
\end{equation}
Hence, the \PARAFAC decomposes the given tensor into a sum of rank one tensors
(see \cite{linalgtensor09, tree-tucker}), which is illustrated by Figure \ref{fig:ParaFac}.
If we think of $\vec x$ as a vector this is equivalent to
\begin{equation*}
\vec x =\sum_{s=1}^D \vec{a_{1}^{(s)}} \otimes \dots \otimes \vec{a_{p}^{(s)}}
\end{equation*}
with tensor products of smaller vectors.
One often finds the normalized form
\begin{equation}\label{eq:ParaFacNormalized}
\vec x =\sum_{s=1}^D \lambda_s \vec{a_{1}^{(s)}} \otimes \dots \otimes \vec{a_{p}^{(s)}}
\end{equation}
with vectors $\vec{a_i^{(s)}}$ of norm~$1$.

If $D$ is minimal in the representation \myref{eq:canonicalDecomposition} of $\vec x$,
it is called the \defini{canonical rank}.

\subsection*{Tensor Train Schemes}
\begin{figure}[ht]
\centering
\subfigure[The Tucker decomposition scheme.]{
\includegraphics[width=0.95\textwidth]{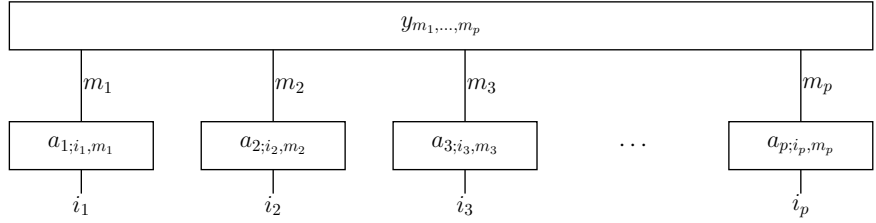}
\label{fig:Tucker}
}
\vfill
\subfigure[The canonical decomposition scheme.]{
\includegraphics[width=0.95\textwidth]{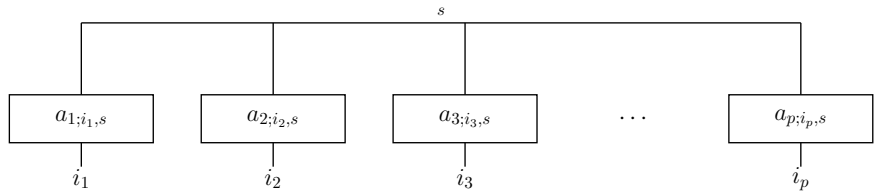}
\label{fig:ParaFac}
}
\vfill
\subfigure[The Tensor Train decomposition scheme]{
\includegraphics[width=0.95\textwidth]{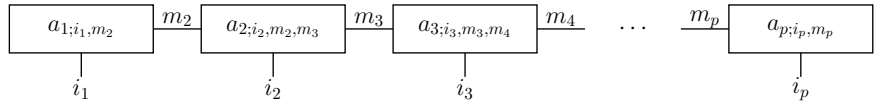}
\label{fig:TensorTrain}
}
\caption{Tensor decomposition schemes}
\label{fig:tensorDecomp}
\end{figure}

Application of the concepts \myref{eq:Tucker} or \myref{eq:canonicalDecomposition} would be
a generalization of SVD that allows a substantial reduction
in number of parameters for deriving a good approximation for a given tensor.
Unfortunately, these decompositions have disadvantages like still exponential growth, lack of
robust algorithms for rank reduction.
Oseledets and Tyrtyshnikov (\cite{OseTyr09Recursive}) proposed the following Tensor Train scheme (\TT)
as an attempt to overcome these difficulties.
In a first step a dyadic decomposition is introduced by cutting the index set and the tensor
in two parts introducing an additional summation with a newly introduced ancilla index $m$:
\begin{equation}
x_{i_1,\dots, i_k;i_{k+1}, \dots, i_p}=\sum_{m} a_{1;i_1,\dots ,i_k,m}\cdot a_{2;i_{k+1},\dots ,i_p,m} \; .
\label{eq:tt1}
\end{equation}
This is the first step of the Tree Tucker \cite{tree-tucker} decomposition.
Now we may apply this process recursively to $a_{1}$ and $a_{2}$.
If we always use the index partitioning $i_j; i_{j+1},\dots,i_p$,
we arrive at the \TT format
\begin{equation}\label{eq:TensorTrain}
x_{i_1,\dots,i_p}=\sum_{m_2,\dots,m_p} a_{1;i_1,m_2} \cdot a_{2;i_2,m_2,m_3} \cdot \dots \cdot a_{p;i_p,m_p} \; .
\end{equation}
Again we can distinguish between given, physical indices $i_j$ and ancilla indices.
Figure \ref{fig:TensorTrain} illustrates the \TT decomposition concept.

The \TT scheme \myref{eq:TensorTrain} is exactly the \MPS form (\ref{eq:mps}) for open boundary conditions:
\begin{equation}
x_{i_1,\dots, i_p}= G_1^{(i_1)} G_2^{(i_2)} \cdot \dots \cdot G_p^{(i_p)}
\end{equation}
with matrices $G_j^{(i_j)}$ of size $D_j \times D_{j+1}$, ($D_1=D_{p+1}=1$),
where the matrix sizes differ and are called the \defini{compression rank}.

In \cite{Khoromskij11ApproximationTC}, Khoromskij generalizes the \TT concept to \defini{Tensor Chains}
via the definition
\begin{equation}\label{eq:TC}
x_{i_1,\dots, i_p} = \sum\limits_{m_1,\dots,m_p}{a_{1;i_1,m_1,m_2} a_{2;i_2,m_2,m_3} \cdot \cdots \cdot a_{p;i_p,m_p,m_1}} \; ,
\end{equation}
which corresponds to \MPS with periodic boundary conditions, i.e.
\begin{equation}
x_{i_1,\dots, i_p}= \trace \left( G_1^{(i_1)} G_2^{(i_2)} \cdot \dots \cdot G_p^{(i_p)} \right) \; .
\end{equation}

Starting with Formula (\ref{eq:tt1}) this step can also be applied recursively to the two newly
introduced tensors $(a_1)$ and $(a_2)$ with any cutting point.
Cutting a tensor in two parts should introduce as many ancilla indices as
described by the tensor network connecting the two parts.
So in $1$D there is only one neighboring connection that is broken up and introduces $1$ additional index.

In the 2D network described in Figure \ref{fig:peps1} we could apply this method by cutting in
a first step the down left knot $i_{11}$, thus introducing two ancilla indices labeled with
$(i_{1,1},i_{1,2})$, resp. $(i_{1,1},i_{2,1})$, representing the broken connections to the neighbors of $i_{1,1}$.
Proceeding in this way knot by knot leads to the \PEPS form.
But generalizing the approach we can also consider general cuts of region $R$
of the tensor network in two regions $R_1$ and $R_2$, replacing the given tensor by a sum over a tensor product
of two smaller tensors with indices related to $R_1$, resp. $R_2$, introducing as many ancilla indices as broken
connections in the cut.

\section{Modifications of \ParaFac and \MPS Representations}\label{sec:generalization}
In this section we want to introduce some own ideas concerning the generalization and modification of both
the \ParaFac decomposition scheme (see \ref{subsec:DecompTensors}) and the \MPS (\TT) format.
These newly presented formats will be used as vector ansatz
for the minimization of the Rayleigh quotient~\myref{eq:minimizeRayleighU}.
As we have pointed out in Section \ref{subsec:CompGroundStatesNumericalAspects},
we are looking for representation schemes that allow
both proper approximation properties
and efficient computations such as fast contractions for inner product calculations
and cheap matrix-vector evaluations.
In view of these requirements
we present both theoretical results concerning the computational complexity
and numerical results showing the benefit of our concepts.

At this point we want to emphasize that our problem does not consist in the decomposition
of a known tensor but to find an appropriate ansatz
for vectors and to formulate algorithms for the ground state computation working on such vector representations.
Hence, in this context we focus on algorithmic considerations and investigate, which modifications and generalizations of the presented concepts are still affordable to solve physical problems,
where we a priori work with approximations with ranks which are given by the physicists.
It turns out that the usage of low-rank approaches suffices to give proper results, compare, e.g., \cite{Verstraete06}.

\subsection{\ParaFac Formats}
\label{subsec:ParaFacFormats}
Any state $\vec{x} \in \mathbb C^{2^p}$ of a physical system can be tensorized artificially in several ways.
In the easiest way we may rewrite $\vec x$ as a $p$th ordered binary tensor of the form
\begin{equation}\label{eq:TensorizationBinary}
 \vec{x} = (x_{i_1, \dots, i_p})_{ i_{\ell} = 0,1} \; .
\end{equation}
But we may also define blockings of larger size,
which will follow the interactions of the physical system (e.g. nearest-neighbor interaction).
These blocking concepts introduce formats with a larger number of degree of freedoms
promising better approximation properties
but still allow efficient computations of matrix-vector and inner products.
For reproducing the physical structure such as internal and external interactions
we will also allow formats with different overlapping blockings in the addends,
compare Subsection \ref{subsec:ModBlockParaFac}.

Such blockings of indices can be seen as tensorizations of lower order $q \le p$:
\begin{equation}\label{eq:TensorizationOfVector}
\vec{x} = (x_{(i_1, \dots, i_{s_1}),\dots,(i_{s_{q-1}+1},\dots,i_{s_q})}) =
(x_{j_1,\dots,j_q}) \; .
\end{equation}
where the $k$th mode combines $t_k := s_k - s_{k-1}$ binary indices and has therefore
index $j_k = (i_{s_{k-1}+1},\dots,i_{s_k})$ of size $2^{t_k}$
(for reasons of consistency we define $s_0=0$ and $s_q=p$).

One way to find a convenient representation is to consider appropriate decompositions of
the tensorization~\myref{eq:TensorizationOfVector}.
In this context we don't consider the Tucker format \myref{eq:Tucker}
as it is only meaningful to decompose very large mode sizes (meaning large sets of blocked indices).

\begin{figure}[htb]
\begin{center}
\includegraphics[width=0.8\textwidth]{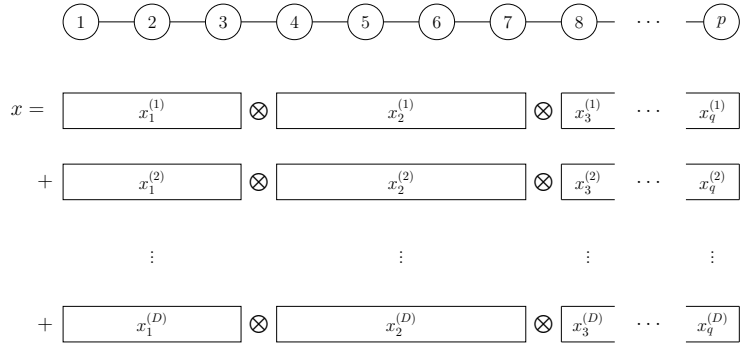}
\end{center}
\caption{\ParaFac ansatz for a chosen tensorization of the vector $\vec{x}$ to be represented.}
\label{fig:ParaFacBlock}
\end{figure}

Hence, from now on we consider the \parafac concept and choose the ansatz vector
to be a sum of rank-$1$ tensors (see Figure \ref{fig:ParaFacBlock}).
In the simplest case \myref{eq:TensorizationBinary}, the \ParaFac decomposition takes the form
\begin{equation}\label{eq:ParaFacBinary}
\vec x = \sum_{s=1}^D \vec{x_1^{(s)}} \otimes \dots \otimes \vec{x_p^{(s)}}
            = \sum_{s=1}^D \mat{ \alpha_1^{(s)} \\ \beta_1^{(s)} } \otimes \cdots \otimes \mat{ \alpha_p^{(s)} \\ \beta_p^{(s)} }  \; ,
\end{equation}
a sum of tensor products of length $2$-vectors.
In view of \myref{eq:MPSasTensorProduct}, the decomposition~\myref{eq:ParaFacBinary} can be seen as a special \MPS form.
Indeed, every \ParaFac representation (\ref{eq:ParaFacBinary}) with $D$ addends corresponds
to an \MPS term with $D \times D$ diagonal matrices.
This fact becomes clear from the construction
$$ A_r^{(0)} = \left(
                 \begin{array}{ccc}
                   \alpha_r^{(1)} &  &  \\
                    & \ddots &  \\
                    &  & \alpha_r^{(D)} \\
                 \end{array}
               \right) \; , \quad A_r^{(1)} = \left(
                                                \begin{array}{ccc}
                                                  \beta_r^{(1)} &  &  \\
                                                   & \ddots &  \\
                                                   &  & \beta_r^{(D)} \\
                                                \end{array}
                                              \right) \; .
 $$

More generally, the \ParaFac scheme for the tensorization~\myref{eq:TensorizationOfVector} leads to the ansatz
\begin{equation}\label{eq:blockParaFac}
\vec x = \sum_{\ell=1}^D \vec{x_1^{(\ell)}} \otimes \dots \otimes \vec{x_q^{(\ell)}} \; .
\end{equation}
with vectors $\vec{x_j^{(\ell)}}$ of moderate size $2^{ t_{j} }$.
Figure \ref{fig:ParaFacBlock} illustrates such a decomposition.

\subsubsection*{Computational Costs}
Inner product calculations of two representations of the form \myref{eq:blockParaFac}
reduce to the inner product of the block vectors:
\begin{equation}\label{eq:innerProductParaFac}
\vec{y}^{\herm} \vec{x} =
\sum_{\ell, \ell'=1}^{D} (\vec{y_1^{(\ell)}}^{\herm} \vec{x_1^{(\ell')}}) (\vec{y_2^{(\ell)}}^{\herm} \vec{x_2^{(\ell')}}) \cdots (\vec{y_q^{(\ell)}}^{\herm} \vec{x_q^{(\ell')}}) \; .
\end{equation}
Therefore, the total costs for each of the $D^2$ inner products are
$$2 (2^{t_1}+2^{t_2}+\cdots + 2^{t_{q}}) + q \leq q(2 l + 1) \leq p (2 l + 1) $$
where $t:=\max\{t_1,t_2,\dots,t_q \}$ is the maximum number
of grouped indices and thus $l=2^t$ denotes the largest block size in $\vec x$.
If all $q$ index sets have the same size $t$ (i.e. $t=p/q$),
the costs can be bounded by $q (2 \cdot 2^{p/q}+1)$.

To compute the matrix-vector product efficiently,
we group the Hamiltonian~$H$ \myref{eq:Hamiltonian} in the same way, i.e.
\begin{equation*}
\begin{split}
H & = \sum_{k=1}^{M} \alpha_k \bigl( Q_1^{(k)} \otimes \dots \otimes Q_{s_1}^{(k)} \bigr) \otimes \cdots \otimes
                                \bigl(Q_{s_{q-1}+1}^{(k)} \otimes \cdots \otimes Q_p^{(k)} \bigr)  \\
  & = \sum_{k=1}^{M} \alpha_k H_1^{(k)} \otimes \cdots \otimes H_{q}^{(k)} \; .
\end{split}
\end{equation*}
For the matrix-vector product we thus obtain
\begin{equation}\label{eq:HamTimesParaFacVector}
\begin{split}
H \vec{x} & = \left( \sum_{k=1}^{M} \alpha_k H_1^{(k)} \otimes \cdots \otimes H_{q}^{(k)} \right)
               \left( \sum_{\ell=1}^D \vec{x_1^{(\ell)}} \otimes \dots \otimes \vec{x_q^{(\ell)}} \right) \\
        & = \sum_{k=1}^{M} \sum_{\ell=1}^D \left( \alpha_k H_1^{(k)} \vec{x_1^{(\ell)}} \right) \otimes
                    \cdots \otimes \left( H_q^{(k)} \vec{x_q^{(\ell)}} \right) \\
        & = \sum_{k=1}^{M} \sum_{\ell=1}^D \vec{y_1^{(k,\ell)}} \otimes \cdots \otimes \vec{y_q^{(k,\ell)}}
                    \; .
\end{split}
\end{equation}
However, typically we do not need the matrix-vector products~\myref{eq:HamTimesParaFacVector} explicitly,
we only require them for inner products of the form $\vec{y}^{\herm} (H \vec{x})$ (compare the nominator of the Rayleigh quotient)
\begin{equation}\label{eq:innerProductHamParaFac}
\vec{y}^{\herm} H \vec{x} = \sum_{k=1}^{M} \alpha_k \sum_{\ell,\ell' =1}^D
                    \left( \vec{y_1^{(\ell')}}^{\herm} H_1^{(k)} \vec{x_1^{(\ell)}} \right)
                        \cdots \left( \vec{y_q^{(\ell')}}^{\herm} H_q^{(k)} \vec{x_q^{(\ell)}} \right) \; .
\end{equation}
The products $H_i^{(k)} \vec{x_i^{(\ell)}}$ can be computed implicitly
without constructing the matrices $H_i^{(k)}$ explicitly.
Each of these small matrix-vector products can be computed linearly
in the size of $H_i^{(k)}$ (i.e. $2^{t_i}$).
Thus, the total costs for each addend in the inner product \myref{eq:innerProductHamParaFac} are
$3 (2^{t_1} + 2^{t_2} + \dots + 2^{t_q}) + q$
and can again be bounded by $q ( 3 \cdot 2^{p/q} + 1)$ in the case of equal block sizes.
Hence, the costs for \myref{eq:innerProductHamParaFac} are $2 M D^2 q (3 \cdot 2^{p/q} + 1)$.

\subsubsection*{Using \ParaFac for the Ground State Problem}
\label{subsec:UsingParaFac}
Let us now apply the \ParaFac approach \myref{eq:blockParaFac}
as ansatz for the Rayleigh quotient minimization.
Then, Eq. \ref{eq:minimizeRayleighU} reads
\begin{equation}
\min \frac{ \left( \sum\limits_{\ell=1}^D \vec{x_1^{(\ell)}} \otimes \dots \otimes \vec{x_q^{(\ell)}} \right)^{\herm} H \left( \sum\limits_{\ell=1}^D \vec{x_1^{(\ell)}} \otimes \dots \otimes \vec{x_q^{(\ell)}} \right) }{ \left( \sum\limits_{\ell=1}^D \vec{x_1^{(\ell)}} \otimes \dots \otimes \vec{x_q^{(\ell)}} \right)^{\herm} \left( \sum\limits_{\ell=1}^D \vec{x_1^{(\ell)}} \otimes \dots \otimes \vec{x_q^{(\ell)}} \right) } \; .
\end{equation}
This minimization task can be realized by an \als-based procedure.

In a first way we could think about the following proceeding:
We start with a \ParaFac representation of rank $1$ ($D=1$),
optimize it via \als and then we successively add one summand and optimize it in an \als-based way.
In the first stage we would have to minimize
\begin{equation*}
\min_{\vec x} \frac{ \vec x^{\herm} H \vec x}{\vec x^{\herm} \vec x} =
\min_{\vec{x_1},\dots,\vec{x_q}}
\frac
{ \left( \vec{x_1} \otimes \vec{x_2} \otimes \dots \otimes \vec{x_q} \right)^{\herm}
H \left( \vec{x_1} \otimes \vec{x_2} \otimes \dots \otimes \vec{x_q} \right)}
{ \left( \vec{x_1} \otimes \vec{x_2} \otimes \dots \otimes \vec{x_q} \right)^{\herm}
  \left( \vec{x_1} \otimes \vec{x_2} \otimes \dots \otimes \vec{x_q} \right)} \; .
\end{equation*}
This optimization problem can be solved via \als:
considering all of the $\vec{x_j}$ up to ${\color{red} \vec{x_i}}$ as fixed,
we obtain
 \begin{equation}\label{eq:parafacRQ1addend}
 \min_{{\color{red} \vec{x_i}}}
 \frac{
    \left( \vec{x_1} \otimes \cdots \otimes {\color{red} \vec{x_i }} \otimes \cdots \otimes \vec{x_q} \right)^{\herm}
    H \left( \vec{x_1} \otimes \cdots \otimes {\color{red} \vec{x_i }} \otimes \cdots \otimes \vec{x_q} \right)
    }
    { \left( \vec{x_1} \otimes \cdots \otimes {\color{red} \vec{x_i }} \otimes \cdots \otimes \vec{x_q} \right)^{\herm}
        \left( \vec{x_1} \otimes \cdots \otimes {\color{red} \vec{x_i }} \otimes \cdots \otimes \vec{x_q} \right)
     } \; .
 \end{equation}
Following \myref{eq:innerProductHamParaFac} and \myref{eq:innerProductParaFac},
we may contract all indices up to $i$ and obtain
\begin{equation*}
\begin{split}
& \min_{ \color{red} \vec{x_i} }
    \frac{ \sum\limits_{k=1}^M \alpha_k (\vec{x_1}^{\herm} H_1^{(k)} \vec{x_1}) \cdots ( {\color{red} \vec{x_i}}^{\herm} H_i^{(k)} { \color{red} \vec{x_i}})
            \cdots ( \vec{x_q}^{\herm} H_q^{(k)} \vec{x_q})
    }{ (\vec{x_1}^{\herm} \vec{x_1}) \cdots ( {\color{red} \vec{x_i}}^{\herm} {\color{red} \vec{x_i}}) \cdots (x_q^{\herm} x_q)
    } \\[1.mm]
= & \min_{ \color{red} \vec{x_i} }
\frac{ \sum\limits_{k=1}^M \alpha_k \beta_k {\color{red} \vec{x_i}}^{\herm} H_i^{(k)} { \color{red} x_i}
    }{ \gamma ( {\color{red} \vec{x_i}}^{\herm} {\color{red} \vec{x_i} })
    }
    = \min_{\color{red} \vec{x_i}}
        \frac{ {\color{red} \vec{x_i}}^{\herm} \left(  \sum\limits_{k=1}^M \frac{ \alpha_k \beta_k}{\gamma} H_i^{(k)} \right)  {\color{red} \vec{x_i} }
        }{ {\color{red} \vec{x_i} }^{\herm} {\color{red} \vec{x_i}} }  \; , \\
\end{split}
\end{equation*}
a standard eigenvalue problem for a matrix of size $2^{t_i} \times 2^{t_i}$ that can be solved via classical iterative methods which only require
the computation of matrix vector products.
These products can be executed implicitly without constructing the matrices explicitly.

Now we suppose that we have already optimized $D-1$ addends
in the representation \myref{eq:blockParaFac} and we want to find the next optimal addend.
This means that the ansatz vector is now of the form
\begin{equation*}
\vec x = \underbrace{\vec{x_1} \otimes \dots \otimes \vec{x_q}}_{= \bigotimes\limits_{j=1}^q \vec{x_j} }
+ \sum_{\ell=1}^{D-1} \underbrace{\vec{y_1^{(\ell)}} \otimes \dots \otimes \vec{y_q^{(\ell)}}}_{=:\vec{y^{(\ell)}}} \; . \end{equation*}
with already optimized $\vec{y_j^{(\ell)}}$-terms and vectors $\vec{x_i}$
that have to be optimized via \ALS.
Contracting over all terms up to ${\color{red} \vec{x_i}}$ (compare Eq. \ref{eq:innerProductHamParaFac}) we obtain
\begin{equation*}
\vec{x}^{\herm} H \vec{x} = {\color{red} \vec{x_i}}^{\herm} H_i {\color{red} \vec{x_i}} +  \vec{u_i}^{\herm} {\color{red} \vec{x_i}} + {\color{red} \vec{x_i}}^{\herm} \vec{u_i} + \beta \; ,
\end{equation*}
where $\vec{u_i}$ and $\beta$ comprise the contractions with the $\vec{y_j}$ terms.
For the denominator we analogously obtain
\begin{equation*}
\vec{x}^{\herm} \vec{x} = {\color{red} \vec{x_i}}^{\herm} \gamma I {\color{red} \vec{x_i}} +  \vec{v_i}^{\herm} {\color{red} \vec{x_i}} + {\color{red} \vec{x_i}}^{\herm} \vec{v_i} + \rho \; .
\end{equation*}
Altogether we have to solve the generalized eigenvalue problem
\begin{equation}\label{eq:generalEigProblemRQ}
\min_{ \color{red} \vec{x_i}} \frac{ \left(
                    \begin{array}{cc}
                      { \color{red} \vec{x_i}}^{\herm} & 1 \\
                    \end{array}
                  \right) \left(
                            \begin{array}{cc}
                              H_i & \vec{u_i} \\
                              \vec{u_i}^{\herm} & \beta \\
                            \end{array}
                          \right) \left(
                                    \begin{array}{c}
                                      { \color{red} \vec{x_i}} \\
                                      1 \\
                                    \end{array}
                                  \right)
}{ \left(
     \begin{array}{cc}
       { \color{red} \vec{x_i}}^{\herm} & 1 \\
     \end{array}
   \right) \left(
             \begin{array}{cc}
               \gamma I & \vec{v_i} \\
               \vec{v_i}^{\herm} & \rho \\
             \end{array}
           \right) \left(
                     \begin{array}{c}
                       { \color{red} \vec{x_i}} \\
                       1 \\
                     \end{array}
                   \right)
}
\end{equation}
It turns out that the denominator matrix can be factorized via a Cholesky factorization.
Therefore we have to solve a standard eigenvalue problem of moderate size.

However, the proposed procedure may cause problems which result from the fact that,
for general tensors, the best rank-$D$ approximation does not have to comprise
the best rank-$(D-1)$ approximation, see, e.g., \cite{TensorReview}.
Our numerical results (Figure \ref{fig:ParaFacNumeric}) will approve this fact.

For this reason we now consider a technique where we use \als to optimize
all the blocks related to the same mode ${\color{red} i}$ simultaneously.
The minimization task takes the form
$$ \min_{\color{red} \vec{x_i}} \frac{
                            \left( \sum\limits_{\ell=1}^D \vec{x_1^{(\ell)}} \otimes \dots \otimes {\color{red} \vec{ x_i^{(\ell)}}}
                                    \otimes \dots \otimes \vec{ x_q^{(\ell)}} \right)^{\herm}
                            H \left( \sum\limits_{\ell=1}^D \vec{x_1^{(\ell)}} \otimes \dots \otimes {\color{red} \vec{x_i^{(\ell)}}}
                                    \otimes \dots \otimes \vec{x_q^{(\ell)}} \right)
                            }{ \left( \sum\limits_{\ell=1}^D \vec{x_1^{(\ell)}} \otimes \dots \otimes {\color{red} \vec{x_i^{(\ell)}}}
                                    \otimes \dots \otimes \vec{x_q^{(\ell)}} \right)^{\herm}
                                \left( \sum\limits_{\ell=1}^D \vec{x_1^{(\ell)}} \otimes \dots \otimes {\color{red} \vec{x_i^{(\ell)}}}
                                    \otimes \dots \otimes \vec{x_q^{(\ell)}} \right) } \; . $$
Contracting all terms in the nominator (compare Eq. \ref{eq:innerProductHamParaFac})
up to the ${\color{red} \vec{x_i^{(\ell)}}}$ vectors results in
\begin{equation*}
\begin{split}
& \sum_{k=1}^{M} \alpha_k \sum_{\ell,\ell' =1}^D
                    \left( \vec{x_1^{(\ell')}}^{\herm} H_1^{(k)} \vec{x_1^{(\ell)}} \right)
                        \cdots \left( \vec{\color{red}x_i^{(\ell')}}^{\herm} H_i^{(k)} {\color{red}\vec{x_i^{(\ell)}}} \right) \cdots
                    \left( \vec{x_q^{(\ell')}}^{\herm} H_q^{(k)} \vec{x_q^{(\ell)}} \right) \\
= & \sum_{\ell,\ell' =1}^D \vec{\color{red}x_i^{(\ell')}}^{\herm} \tilde H_i^{(\ell',\ell)} {\color{red}\vec{x_i^{(\ell)}}} \; .
\end{split}
\end{equation*}
For the denominator, we analogously obtain
\begin{equation*}
\sum_{\ell,\ell' =1}^D \vec{\color{red}x_i^{(\ell')}}^{\herm} \left( \beta_{\ell',\ell} I \right) {\color{red}\vec{x_i^{(\ell)}}} \; .
\end{equation*}
Altogether, the approach leads to the minimization problem
\begin{equation}\label{eq:SimParaFacGenEVP}
\min_{\color{red} \vec{x_i}}
\frac{
    \mat{ \bigl({\color{red} \vec{x_i^{(1)  }} }\bigr)^{\Herm} , \dots , \bigl({\color{red} \vec{x_i^{(D)  }} }\bigr)^{\Herm} }
    \mat{ \tilde H_i^{(1,1)} & \dots & \tilde H_i^{(1,D)} \\ \vdots & \ddots & \vdots \\ \tilde H_i^{(D,1)} & \dots & \tilde H_i^{(D,D)}}
    \mat{ {\color{red} \vec{x_i^{(1) }} } \\ \vdots \\ {\color{red} \vec{x_i^{(D) }} } }
}{
    \mat{ \bigl({\color{red} \vec{x_i^{(1)  }} }\bigr)^{\Herm} , \dots , \bigl({\color{red} \vec{x_i^{(D)  }} }\bigr)^{\Herm} }
    \mat{ \beta_{1,1} I & \dots & \beta_{1,D} I \\ \vdots & \ddots & \vdots \\ \beta_{D,1} I & \dots & \beta_{D,D} I}
    \mat{ {\color{red} \vec{x_i^{(1) }} } \\ \vdots \\ {\color{red} \vec{x_i^{(D) }} } }
} \; ,
\end{equation}
a generalized eigenvalue problem of size $ D 2^{t_i} \times D 2^{t_i}$.
This approach requires the solution of larger eigenproblems, but by increasing the set of variables
to be optimized it turns out that we require less optimization steps
and obtain better convergence results (see Figure \ref{fig:ParaFacNumeric}).

\subsubsection*{Formulation of an Algorithm}
\begin{algorithm}[htbp]
\caption{Rayleigh Quotient Minimization in the \ParaFac model}
\label{alg:simultaneousParaFacMinimization}
\begin{algorithmic}[1]
\STATE Provide initial guess for all vectors $\vec{x_i^{(\ell)}}$ ($i=1,\dots,q$ and $\ell=1,\dots,D$)
\WHILE{Not yet converged}
\FOR{$i=1,\dots,q$}
\STATE Compute all contractions in $\vec{x}^{\herm} H \vec{x}$ and $\vec{x}^{\herm} \vec{x}$ up to index $i$
\STATE Solve the generalized eigenvalue problem \myref{eq:SimParaFacGenEVP}
\ENDFOR
\STATE Normalize each addend in the \ParaFac representation (\ref{eq:ParaFacNormalized})
\ENDWHILE
\end{algorithmic}
\end{algorithm}
As possibility to choose the initial guesses, we propose to set
$ \{ \vec{x_i^{(1)}}, \dots , \vec{x_i^{(D)}} \}$ to $D$ linearly independent eigenvectors of
$$ H_i = \sum_{k=1}^M Q_{s_{i-1}+1}^{(k)} \otimes \cdots \otimes Q_{s_i}^{(k)} \; .$$
As stopping criterion we may define a maximum number of iterations or
we could specify a stopping criterion based on the degree of improvement.
One possibility to accelerate the \als-based optimization procedure
is to apply the Enhanced Linesearch method \cite{ELS08}, which is not considered here.

\subsubsection*{Numerical Results}
\begin{figure}[ht]
\centering
\subfigure[$10$ spins, one-by-one optimization.]{
\includegraphics[width=0.46\textwidth]{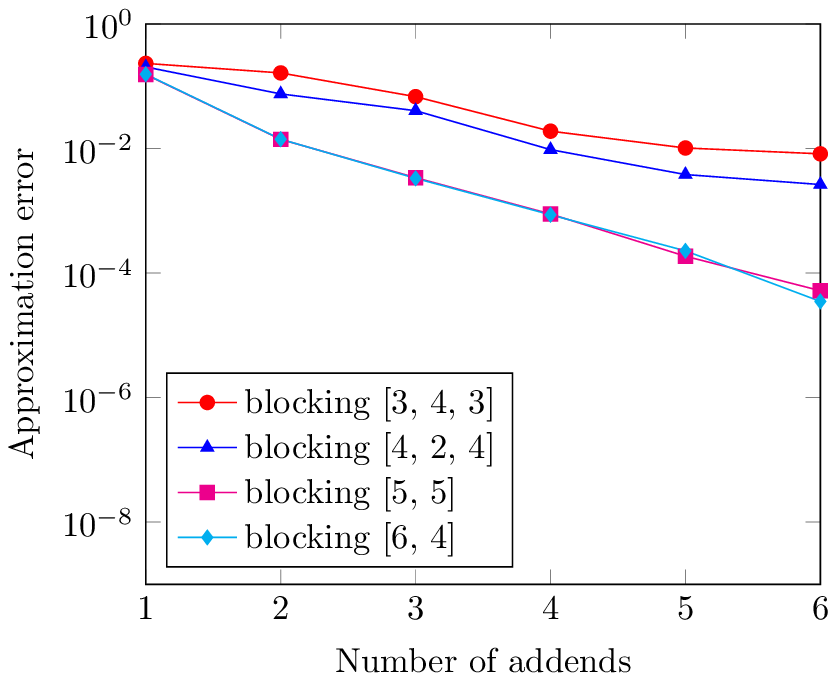}
\label{fig:ParaFacNumericp10}
}
\subfigure[$10$ spins, simultaneous optimization.]{
\includegraphics[width=0.46\textwidth]{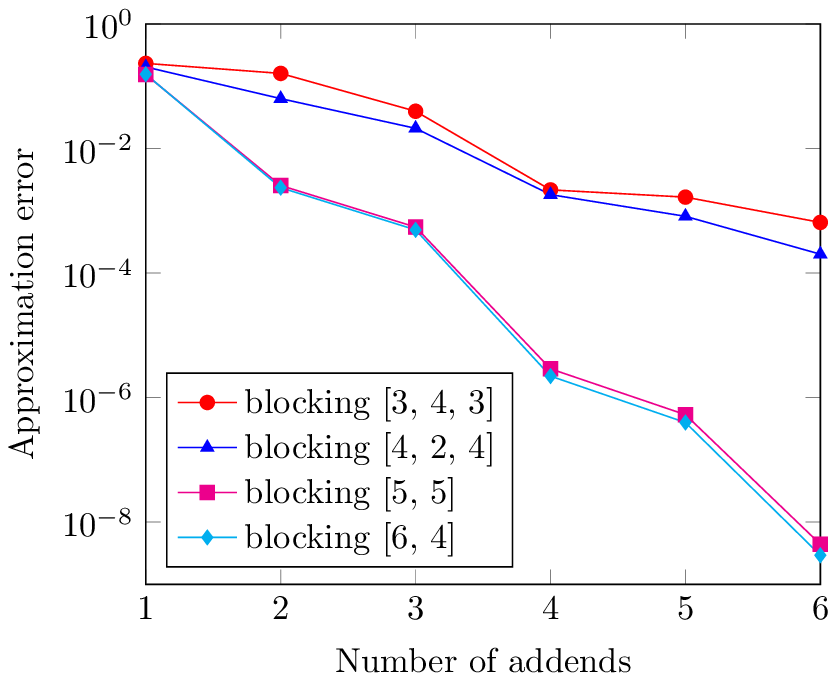}
\label{fig:ParaFacNumericp10Simultan}
}
\subfigure[$12$ spins, one-by-one optimization.]{
\includegraphics[width=0.46\textwidth]{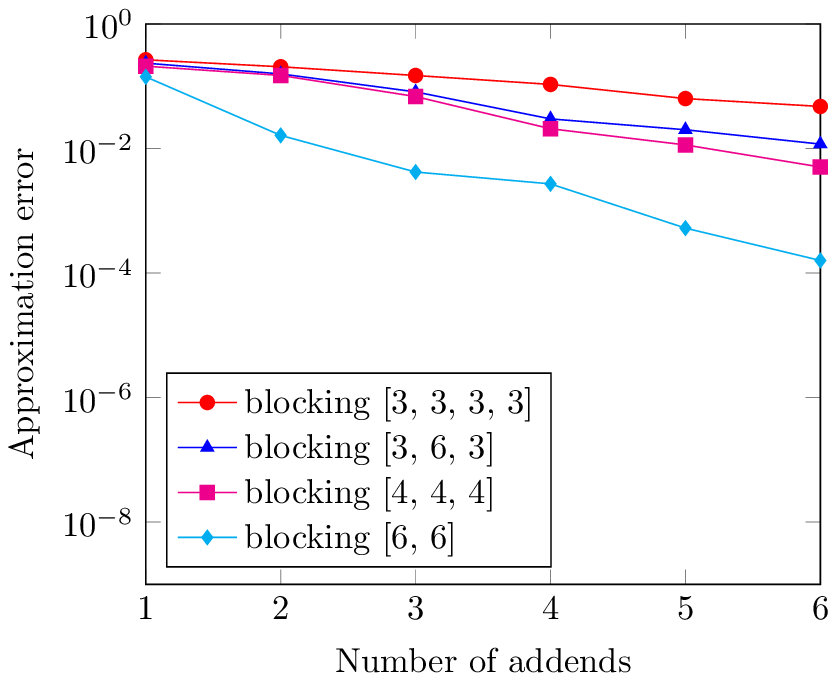}
\label{fig:ParaFacNumericp12}
}
\subfigure[$12$ spins, simultaneous optimization.]{
\includegraphics[width=0.46\textwidth]{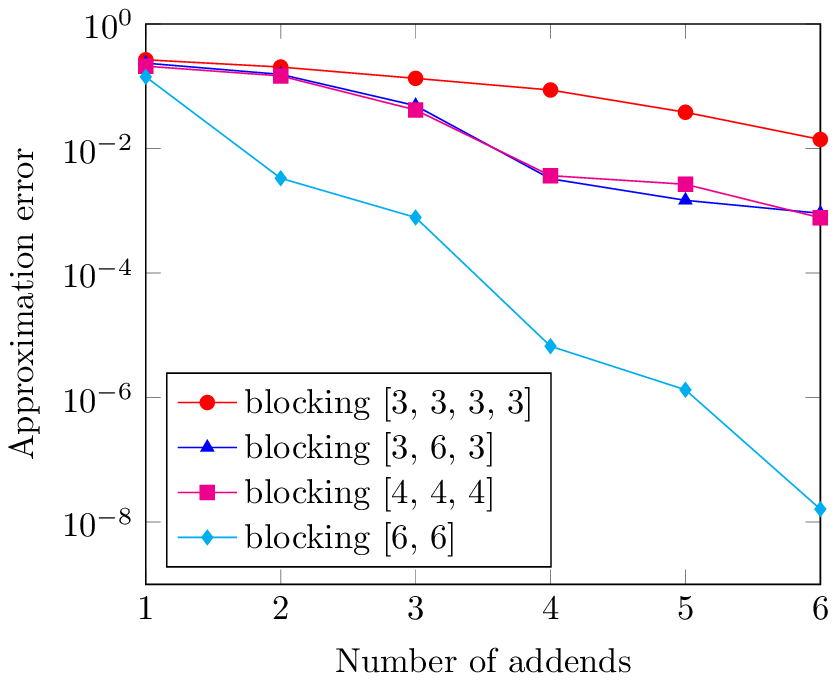}
\label{fig:ParaFacNumericp12Simultan}
}
\caption{Approximation error for the computation of the ground state energy of a $10$ spin and $12$ spin
Ising-type Hamiltonian.}
\label{fig:ParaFacNumeric}
\end{figure}

After these technical derivations, we want to show first numerical results.
We computed the ground state of a $10$ spin and a $12$ spin Ising-type Hamiltonian~(\ref{eq:IsingModel})
using different blocking schemes $[t_1,\dots,t_q]$.
The \als-based Rayleigh quotient minimization was performed in two different ways:
a one-by-one minimization following \myref{eq:generalEigProblemRQ} and a simultaneous minimization
described by Algorithm~\ref{alg:simultaneousParaFacMinimization}.
It turns out that the results are getting better when using larger vector blocks.
This consideration is based on the fact that using larger blocks leads to a larger number of degrees of freedom.
Figure \ref{fig:ParaFacNumeric} depicts the effect of using different \ParaFac representations
and shows the benefit resulting from the simultaneous optimization.

\subsection{Block MPS}\label{subsec:blockmps}
In the block \MPS ansatz we group single particles together and apply \MPS concepts to these grouped blocks.
The \MPS ansatz (\ref{eq:mps}) is based on using $2$ matrices $A_j^{(i_j)}$ at position $j$.
A first generalization of \MPS vectors can be derived by blocking indices in subgroups.
So we can combine indices $(i_1,...,i_r)$ to index $j_1$, and indices $(i_{r+1},...,i_p)$ to index $j_2$.
Then the \MPS vector to this blocking with open boundary conditions would give
\begin{equation*}
x_i=x_{i_1,...,i_p}= x_{ \{i_1,\dots,i_r\},\{i_{r+1},\dots,i_p\}} = x_{j_1, j_2} =
{A_1^{(j_1)}} \cdot A_2^{(j_2)}
\end{equation*}
or vector
\begin{equation}\label{eq:mpscoarse2terms}
\vec x = \sum_{m_2=1}^D \vec{a_{1;1,m_2}} \otimes \vec{a_{2;m_2,1}}
\end{equation}
with $D$ pairs of short vectors of length $2^r$, resp. $2^{p-r}$.
Compared to standard \MPS we have different numbers for the degree of freedom (DOF), and different
costs for inner products.
\MPS combines in each term $p$ vectors of length $2$ in a tensor product and combines $D^p$ vectors;
thus, it has $2pD^2$ DOF's and $2pD^3$ costs for inner products.
The $2$-term block coarse grain \MPS form \myref{eq:mpscoarse2terms} combines in each term $2$ vectors of total length $2^r$ and $2^{p-r}$.
Therefore, it has $D(2^r+2^{p-r})$ DOF's and the estimate for the inner product costs is $D^2(2^r+2^{p-r})$.
The computation of the full inner product reduces to $D^2$ inner products of short vectors at
position $1$, resp. $2$:
\begin{equation*}
\begin{split}
\vec x^{\herm} \vec x & = \sum_{m_2'}^D \vec{a_{1;1,m_2'}}^{\herm}\otimes \vec{a_{2;m_2',1}}^{\herm}\cdot
        \sum_{m_2}^D \vec{a_{1;1,m_2}}\otimes \vec{a_{2;m_2,1}}  \\
      &  = \sum_{m_2,m_2'} (\vec{a_{1;1,m_2'}}^{\herm} \vec{a_{1;1,m_2}})\cdot (\vec{a_{2;m_2',1}}^{\herm} \vec{a_{2;m_2,1}}) \;.
\end{split}
\end{equation*}

For using three blocks we derive
\begin{equation*}
x_i=x_{ \{i_1,\dots,i_{r_1}\}, \{i_{r_1+1},\dots, i_{r_2} \}, \{ i_{r_2 + 1}, \dots i_p \} }=x_{j_1,j_2,j_3}={A_1^{(j_1)}} A_2^{(j_2)} A_3^{(j_3)}
\end{equation*}
or vector
\begin{equation*}
\vec x=\sum_{m_2,m_3}^D \vec{a_{1;1,m_2}}\otimes \vec{a_{2;m_2,m_3}}\otimes \vec{a_{3;m_3,1}}
\end{equation*}
with $D 2^{t_1}+ D^2 2^{t_2}+ D 2^{t_3}$ DOF's, $t_1 = r_1, t_2 = r_2 -r_1, t_3 = p-r_2$,
combining $D^2$ long vectors,
and costs of order $D^4$ for an inner product in view of
\begin{equation*}
\vec x^{\herm} \vec x=
\sum_{m_2,m_3 \atop m_2',m_3'}
({\vec{a_{1;1,m_2'}}^{\herm}}\vec{a_{1;1,m_2}})\cdot ( {\vec{a_{2;m_2',m_3'}}^{\herm}} \vec{a_{2;m_2,m_3}})\cdot
(\vec{a_{3;m_3',1}}^{\herm} \vec{a_{3,m_3}})\; .
\end{equation*}
By using the \MPS contraction scheme (Figures \ref{fig:mps1} -- \ref{fig:mps6}), the costs for
the inner product can be reduced to $D^3(2^{t_1}+2^{t_2}+2^{t_3})$.

In general, the Block \MPS ansatz with $k$ blocks gives
\begin{equation}\label{eq:blockMPSMatrix}
x_i=x_{i_1,...,i_p}=x_{j_1,\dots ,j_k}={A_1^{(j_1)}}\cdot A_2^{(j_2)}\cdots
 A_{k-1}^{(j_{k-1})} \cdot A_k^{(j_k)}
\end{equation}
or vector
\begin{equation}
\vec x=\sum_{m_2,\cdots ,m_k} \vec{a_{1;1,m_2}} \otimes \vec{a_{2;m_2,m_3}}\otimes \cdots
\vec{a_{k-1;m_{k-1},m_k}}\otimes  \vec{a_{k;m_k,1}} \; .
\end{equation}
DOF's: $D(2^{t_1}+2^{t_k})+D^2(2^{t_2}+\cdots + 2^{t_{k-1}})$ for $t_1 = r_1$, $t_k = r_k - r_{k-1}$.

\noindent Costs: $D^3(2^{t_1}+\cdots + 2^{t_k})$ using the \MPS contraction scheme,

\noindent Combination of $D^{k-1}$ full vectors.

Altogether, the Block \mps ansatz causes greater effort for the computation of inner products,
but the increasing number of degrees of freedom may allow better approximation properties.
The Block \MPS format corresponds to the Tensor Train format with mode sizes $2^{t_i}$, $i=1,\dots,k$.

In summary, \MPS can be considered as a fine grain method, using many vectors, while the coarse grain
\MPS uses only a sum of tensor products of a few large vectors respectively.
The fine grain \MPS uses at each position $s$ a matrix pair $A_s^{(i_s)}$, $i_s=0,1$,
while the $k$-form \myref{eq:blockMPSMatrix} uses $2^{t_s}$ matrices $A_s^{(j_s)}$, $j_s=0,1,...,2^{t_s}-1$.
In the same way the length of vector $\vec{a_{s;m_s,m_{s+1}}}$ is $2^{t_s}$.

\subsection{Mixed Blockings in the \ParaFac Ansatz}\label{subsec:ModBlockParaFac}
\begin{figure}[ht]
\includegraphics[width = 0.85\textwidth]{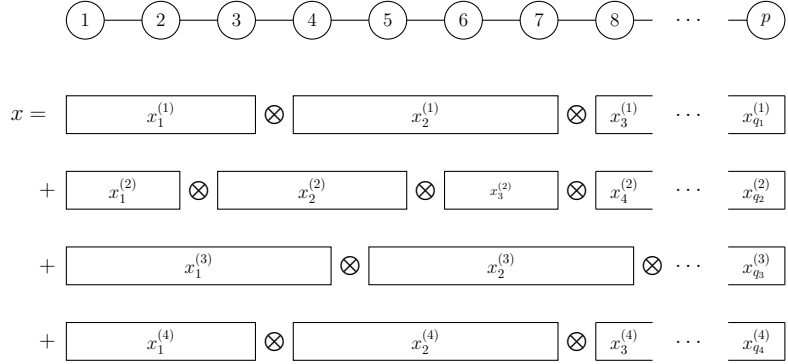}
\caption{A \PARAFAC-like representation.
In contrast to the decomposition scheme illustrated by Figure \ref{fig:ParaFacBlock} different blockings are possible.}
\label{fig:ParaFacMixedBlock}
\end{figure}

In this section, we want to analyze mixed tensor representations similar to \PARAFAC concepts
presented in Section \ref{subsec:UsingParaFac} but allowing different blockings in the addends
(see Figure \ref{fig:ParaFacMixedBlock}).
This may also be interesting for mixing, e.g., Block-\MPS vectors with different block structure
(compare Subsection \ref{subsec:blockMPSMixedBlocking}).

\textbf{Reasoning for mixed blockings:} In connection with PARAFAC or MPS,
approximations of a given vector as a sum of tensor products of smaller vectors
(such as (\ref{eq:MPSasTensorProduct}) or (\ref{eq:blockMPSmixedBlocking}))
which are of different sizes might give different approximation qualities.
E.g. using a partitioning with two vectors is related to a certain matricization of the given vector.
The matricization leads to an SVD with related singular values of certain decay behavior
and therefore a certain truncation error
depending on the number of singular values that are used for the approximation.
Hence, it might be useful to combine different matricizations in \PARAFAC
allowing tensor products of different partitionings.

In the following we want to discuss the question which blockings allow efficient contractions
similar to the case of uniform blockings.
Using such mixed schemes for finding ground states
also leads to generalized eigenproblems
similar to (\ref{eq:generalEigProblemRQ},\ref{eq:SimParaFacGenEVP}),
but now the computation of inner products is generally more costly.
In contrast, the application of the Hamiltonian (\ref{eq:Hamiltonian}) to a mixed vector representation
can again be computed efficiently because the Hamiltonians are given by Kronecker products of $2 \times 2$ matrices.
The question here is, what type of blocking allows fast and efficient contractions in inner products
for vectors with different tensor blocking.

In the following we will consider the problem what tensor aware permutations can be
combined that allow fast and efficient contractions.
Such permutations allow a generalization of \PARAFAC, where in each approximation step
a new tensor aware permutation can be applied before the new term is computed.

So in the first, elementary case we consider
\begin{equation}\label{eq:parafacGeneral}
\begin{split}
\vec{z} = ( z_{i_1,\dots,i_p} )  & = ( x_{1;i_1,\dots,i_{r_1}} ) \otimes \cdots \otimes ( x_{k;i_{r_{k-1}+1}, \dots , i_p}) \\
& \ +  ( y_{1;i_1, \dots,i_{s_1}} ) \otimes \cdots \otimes ( y_{m;i_{s_{m-1}+1},\dots,i_p} )
\end{split}
\end{equation}
where the vector $\vec{z}$ with $2^p$ components is decomposed in two vectors that are Kronecker
products of smaller vectors.
The length of the $k$ vectors $\vec{x_j} = (x_{j,i_{r_{j-1}+1},\dots, i_{r_j}})$ is given by $2^{r_j-r_{j-1}}$
and similarly the $m$ vectors $\vec{y_j}$ are of length $2^{s_j-s_{j-1}}$.
The blocking of $\vec x$ and $\vec y$ may be different.
Such forms can be seen as generalizations of \PARAFAC, or the coarse grain \MPS form.

The interesting question is whether it is possible to introduce efficient contraction
schemes for such sums.
So in the following we want to analyze the costs for inner products of two vectors
that are built as tensor products, but with different block structure as
displayed in Figures \ref{fig:par2} and \ref{fig:par3}.

\subsubsection*{The One-Dimensional Case}
\label{sec:general1D}
If both vectors have the same block structure,
the inner product reduces to the inner product of the block vectors, compare \myref{eq:innerProductParaFac}.
The costs can then be bounded by  $3 k 2^{p/k}$.

\begin{figure}[ht]
\begin{center}
\includegraphics[width=0.95\textwidth]{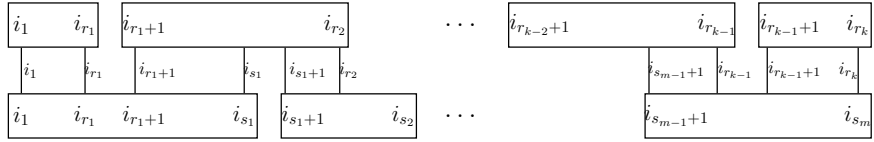}
\end{center}
\caption{Contraction scheme for the calculation of the inner product of two vectors, which are formed by tensor products of smaller vectors, in the open boundary case.}
\label{fig:par2}
\end{figure}

The more interesting case appears if we allow different block sizes.
First, let us consider the open boundary case where the first block in both vectors starts with $i_1$
and the last block ends with $i_p$, compare Figure \ref{fig:par2}.
The contraction starts with the left block computing the summation over $i_1,...,i_{r_1}$:
\begin{equation}\label{eq:ContractDifferentBlocksParaFac}
\sum_{i_1,\dots,i_{r_1}} x_{1;i_1,\dots,i_{r_1}} \cdot y_{1;i_1,\dots, i_{r_1}, \dots, i_{s_1}}
\rightarrow
y_{1;i_{r_1+1},\dots, i_{s_1}} \; .
\end{equation}
This leads to an update of $\vec{ \bar y_1}$ which is replaced by a new vector to indices
$i_{r_1+1},...,i_{s_1}$.
The costs are $2^{r_1}$ for one summation, and the summation has
to be done for all indices $i_{r_1+1},...,i_{s_1}$, hence $2^{s_1-r_1}$ times, leading to
total costs of $2^{s_1}$ for this step.
Now we have to repeat this procedure for the pair
$x_{2;i_{r_1+1},\dots,i_{r_2}}$ and the updated $y_{1;i_{r_1+1}\cdots i_{s_1}}$.
Let us assume that all blocks have size bounded by $2^r$.
In each step the costs are bounded by the size of the longest block, and the block size
of the intermediate generated vectors is smaller than the size of the previous vectors.
Hence in total the costs are bounded by the number of necessary contractions $=k+m$ times
the maximum block size: $ 2^r\cdot (k+m)$.

\begin{figure}[ht]
\begin{center}
\includegraphics[width=0.95\textwidth]{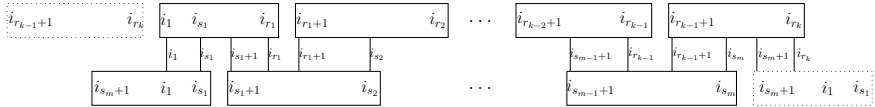}
\end{center}
\caption{Contraction scheme in the periodic boundary case. Differing to open boundary case
illustrated in Figure \ref{fig:par2}, the first and last blocks also overlap.}
\label{fig:par3}
\end{figure}

In the periodic case, the first and the last blocks are allowed to overlap as displayed in
Figure \ref{fig:par3}.
In this case the first contraction looks like
\begin{equation}
\sum_{i_1\cdots i_{s_1}} x_{1;i_1,\dots, i_{s_1}, \dots i_{r_1}}\cdot
   y_{m;i_{s_m+1},\dots, i_p,i_{1},\dots,i_{s_1}}
\rightarrow
z_{1;i_{s_m+1},\dots, i_p, i_{s_1+1}, \dots, i_{r_1}} \; .
\end{equation}
The costs are given by the length of the contraction times the number of times it has to be executed,
that is $2^{s_1}\cdot 2^{p-s_m}\cdot 2^{r_1-s_1}$.
In the worst case the summation part could be only one index which leads to $2^{2r-1}$
costs for maximum block size $2^r$, where
$r:=\max\{r_1,r_2-r_1,\dots, r_k - r_{k-1},s_1, s_2 - s_1, \dots, s_m - s_{m-1}\}$.
But in each step we can choose a contraction where the summation part is larger than the remaining indices.
Then, the costs for one contraction are bounded by $2^{r_1+(p-s_m)}\leq 2^{3r/2}$.
Therefore, in total the costs are bounded by $2^{3r/2}(m+k)$.

\subsection{Mixed Blockings for Block \MPS Vectors}\label{subsec:blockMPSMixedBlocking}
\begin{figure}[ht]
\begin{center}
\includegraphics[width=0.8\textwidth]{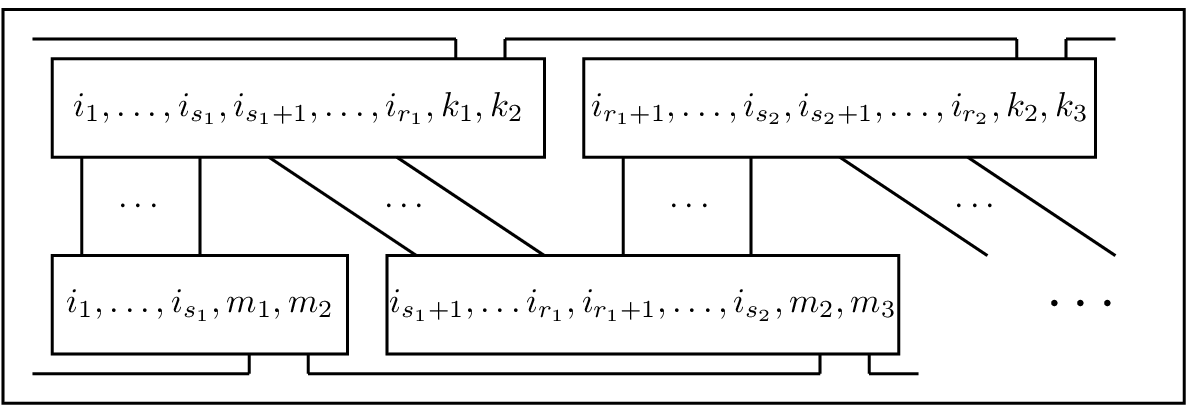}
\end{center}
\caption{Contraction scheme for the inner product of the block \MPS vectors \myref{eq:blockMPSmixedBlocking}.}
\label{fig:contractBlockMPS}
\end{figure}

The contraction schemes \myref{eq:ContractDifferentBlocksParaFac} presented in the previous subsection
may also be applied to calculate inner products of Block \MPS vectors \myref{eq:blockMPSMatrix}
as introduced in Section \ref{subsec:blockmps}.
To this end we have to append the ancilla indices resulting from the matrix products to the physical indices.
For contracting the inner product of the Block \MPS vectors
\begin{equation}\label{eq:blockMPSmixedBlocking}
\begin{split}
\vec x & = \sum\limits_{i_1,\dots,i_p} A_1^{(i_1,\dots,i_{r_1})} A_2^{(i_{r_1 + 1},\dots,i_{r_2})} \cdots A_k^{(i_{r_{k-1}+1},\dots,i_p)} \; ,\\
\vec y & = \sum\limits_{i_1,\dots,i_p} B_1^{(i_1,\dots,i_{s_1})} B_2^{(i_{s_1 + 1},\dots,i_{s_2})} \cdots B_m^{(i_{s_{m-1}+1},\dots,i_p)} \; ,
\end{split}
\end{equation}
we can proceed in the same way as for standard \MPS vectors
(see Figures \ref{fig:contractMPS}, \ref{fig:contractMPS2}),
but instead of contracting over only one physical index $i_j$ per step,
we have to sum over all overlapping indices as illustrated in Figure \ref{fig:contractBlockMPS}.

%

\subsection{The \ParaFac Model for the Two-Dimensional Case}
\begin{figure}[htb]
\begin{center}
\includegraphics[width=0.5\textwidth]{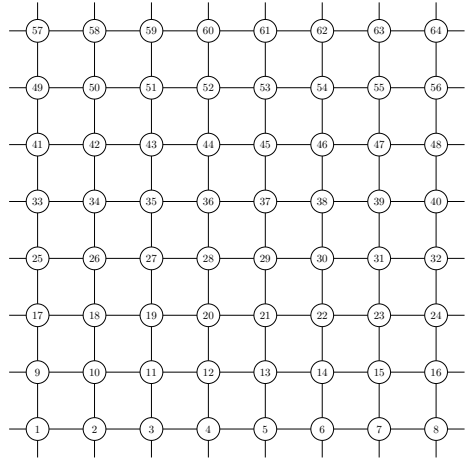}
\end{center}
\caption{A 2D physical system with nearest neighbor interaction in both directions and periodic boundary conditions.}
\label{fig:parpeps1}
\end{figure}

For a 2D tensor network, we introduce another block structure that also allows fast contractions.
As example we consider a 2D physical system with nearest-neighbor interaction in both directions
as displayed in Figure \ref{fig:parpeps1}.

\begin{figure}[ht]
\centering
\subfigure[Grouping of indices.]{
\includegraphics[width=0.45\textwidth]{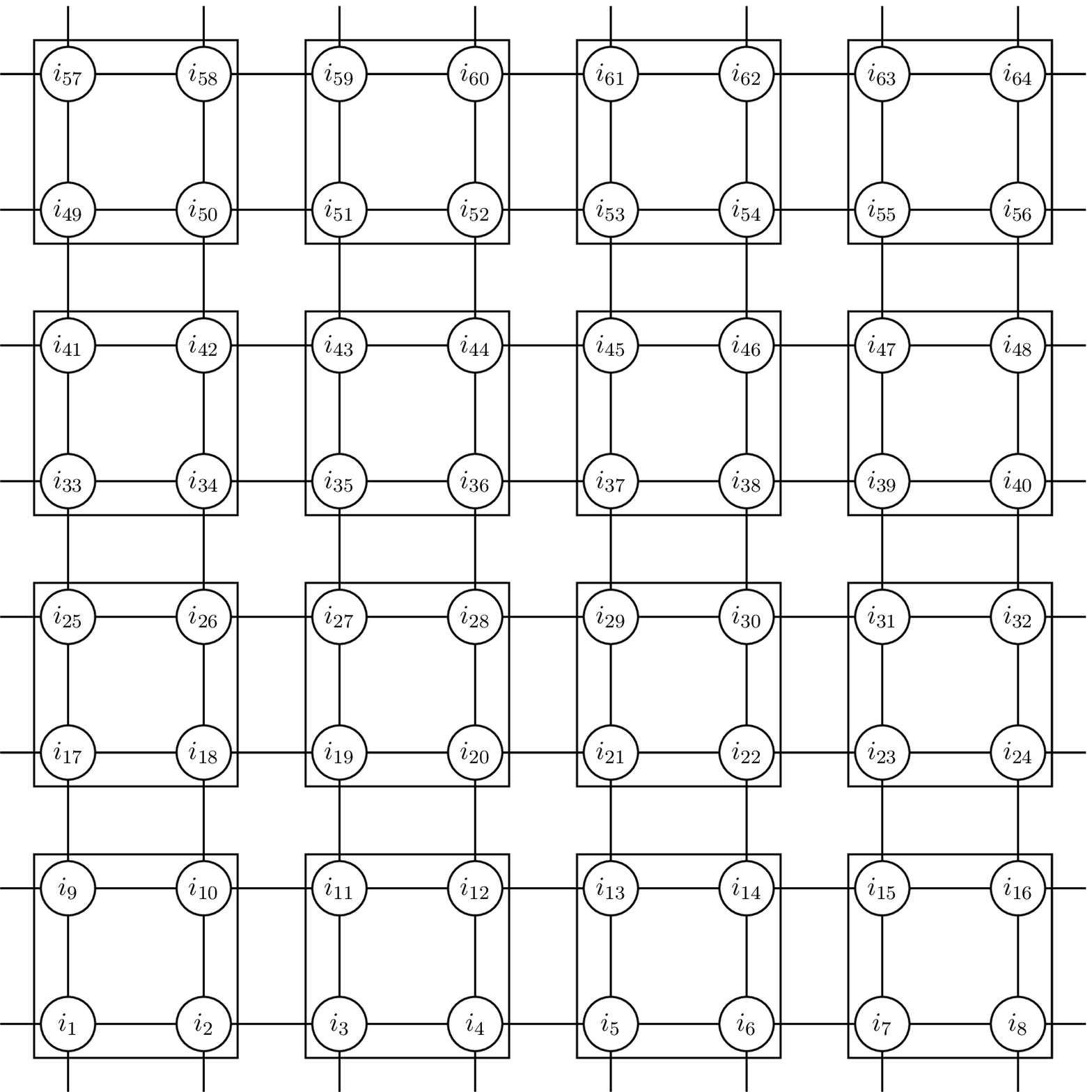}
\label{fig:parpeps2}
}
\subfigure[Introduction of new indices.]{
\includegraphics[width = 0.45\textwidth]{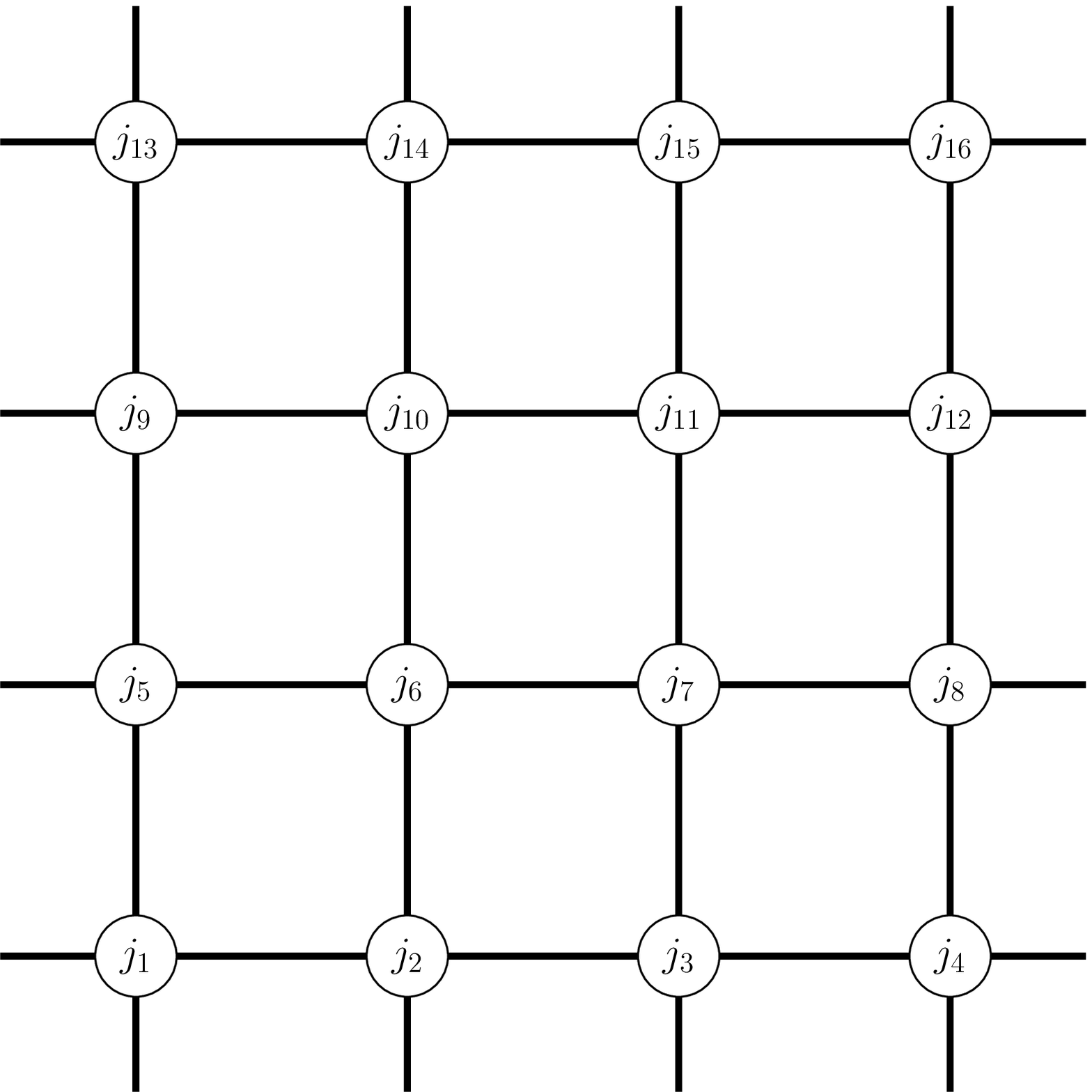}
\label{fig:parpeps3}
}
\caption{The 2D physical system as introduced in Figure \ref{fig:parpeps1},
but now adjacent indices are grouped together and denoted by new indices~$j$.
The original neighborhood relations from Figure \ref{fig:parpeps1} are still taken into account.}
\label{fig:parpeps2+3}
\end{figure}

In a first step we group $r$ indices together to larger
subblocks taking into account the neighborhood relations, see
Figure \ref{fig:parpeps2}.

We denote the newly introduced blocks by indices $j_{s}$.
In the above example each $j$ block contains $4$ physical $i$ indices (Figure \ref{fig:parpeps3}).

\begin{figure}[ht]
\begin{center}
\includegraphics[width=0.95\textwidth]{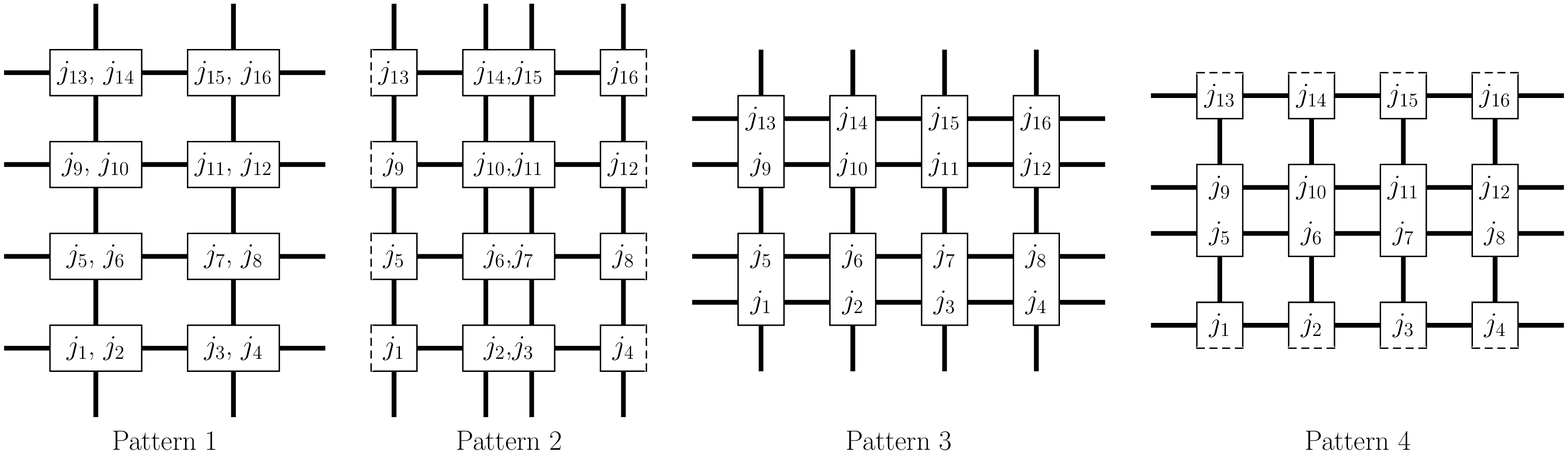}
\end{center}
\caption{Four different patterns of subblocks related to the four vector types defined in \myref{eq:2dim4PatternTypes}}
\label{fig:parpeps5}
\end{figure}

Based on the introduced coarse numbering, we define a final collection of blocks
by building pairs of subblocks.
We allow four different patterns of superblocks (see Figure~\ref{fig:parpeps5}),
which enable efficient contractions in the calculation of mixed inner products.

These four patterns are related to four different vector types
\begin{equation}\label{eq:2dim4PatternTypes}
\begin{split}
\vec{x}^1 & =  \vec{a_{j_1,j_2}} \otimes  \vec{a_{j_3,j_4}} \otimes  \vec{a_{j_5,j_6}}
            \otimes \cdots \otimes \vec{a_{j_{13}, j_{14}}} \otimes \vec{a_{j_{15}, j_{16}}} \; , \\
\vec{x}^2 & =  \vec{a_{j_4,j_1}} \otimes \vec{a_{j_2,j_3}} \otimes  \vec{a_{j_8,j_5}}
            \otimes \cdots \otimes \vec{a_{j_{16}, j_{13}}} \otimes \vec{a_{j_{14},j_{15}}} \; ,\\
\vec{x}^3 & =  \vec{a_{j_1,j_5}} \otimes \vec{a_{j_2,j_6}} \otimes  \vec{a_{j_3,j_7}}
            \otimes \cdots \otimes \vec{a_{j_{11}, j_{15}}} \otimes \vec{a_{j_{12},j_{16}}} \; ,\\
\vec{x}^4 & =  \vec{a_{j_{13}, j_1}} \otimes \vec{a_{j_{14},j_2}} \otimes  \vec{a_{j_{15},j_3}}
            \otimes \cdots \otimes \vec{a_{j_{7}, j_{11}}} \otimes \vec{a_{j_{8},j_{12}}} \; .
\end{split}
\end{equation}

As ansatz vector we may consider
combinations of the four vector patterns:
$$ \vec x = \sum\limits_{\imath} \alpha_{\imath} \vec{x_{\imath}^1}
            + \sum\limits_{\jmath} \beta_{\jmath} \vec{x_{\jmath}^2}
            + \sum\limits_{\ell} \gamma_{\ell} \vec{x_{\ell}^3}
            + \sum\limits_{\kappa} \delta_{\kappa} \vec{x_{\kappa}^4} \; .$$

Contractions between vectors of type $1-4$ can be done in
$\mathcal O(2^{3r})$ operations.
Each contraction considers a block of vector $\vec{x_s}$ combined with a block of vector $\vec{x_t}$
that has an index $j_k$ in common.
Therefore, the costs are the subblocksize to the power of three.
The result of such a contraction is a new block consisting of the pair of indices
that were left over.
Hence the total costs are the total numbers of blocks times the subblocksize to the power of three.

\textbf{Remark:} Similar to the generalization of \MPS to Block \MPS
with mixed blockings as introduced in Section \ref{subsec:blockMPSMixedBlocking},
the mixed blocking formats for 2D systems could also be applied
to modify the \PEPS ansatz (see Section \ref{subsec:peps}).
For efficient computations of inner products we may generalize the contraction scheme (Figures \ref{fig:pepsContraction1} and \ref{fig:pepsContraction2}) to sets of overlapping physical indices.

\section{Conclusions}
\label{sec:conclusions}
This work combines similar concepts proposed in quantum information and linear algebra communities
and presents them from a (multi-) linear algebraic point of view
in a unifying mathematical and computational framework.
In this context we are mainly interested in algorithmic considerations.

For our main task, the computation of ground states of physical quantum systems,
we have to minimize the Rayleigh quotient over an exponentially growing vector space.
To overcome the curse of dimensionality we have to find appropriate formats which allow
cheap computations of matrix products and inner products, but still guarantee proper approximation properties.
These formats are structured in such a way that they enable relaxation methods such as Alternating Least Squares.

Physicists have developed concepts like Matrix Product States, which correspond to Tensor Trains,
for linear one-dimensional problems
or Projected Entangled Pair States for two-dimensional problems.
We present these concepts from a mathematical point of view and show how computations
such as contractions of inner products can be performed efficiently.

As an ansatz format to minimize the Rayleigh quotient we also consider the \ParaFac format, which allows modifications in several directions,
such as different blocking structures.
These generalizations, which can also be applied to modify, e.g., the \MPS or \PEPS ansatz,
are constructed in such a way that
\begin{itemize}
\item[1.] the physical (inter)actions (e.g. nearest neighbor interactions)
can be reproduced properly,
\item[2.] representations are only polynomial in the system size $p$,
\item[3.] computations such as inner product contractions can still be performed efficiently.
\end{itemize}

Our mixed blocking ansatz in \ParaFac and \MPS
is an efficient and flexible way to represent states of physical systems
and can easily be extended to higher dimensions or systems with more complex interactions.
It is, however, an open question, how the difference in the blockings can be constructed automatically
and how efficient they can be for general problems.
\section*{Acknowledgements}
\label{sec:acknowledgements}

This work roots back in a presentation at the $26^{\rm th}$
GAMM Seminar on Tensor Approximations at the Max-Planck-Institute for
Mathematics in the Sciences in Leipzig, February 2010.
It was supported in part by the Bavarian excellence network ENB
via the International Doctorate Program of Excellence
{\em Quantum Computing, Control, and Communication} (QCCC)
as well as by the EU programmes QAP,
\mbox{Q-ESSENCE}, and the exchange with COQUIT. --- We wish to thank Norbert
Schuch for many helpful comments.

\bibliographystyle{elsarticle-num}
\bibliography{compQuantTensorNetworks}

\begin{thebibliography}{10}
\expandafter\ifx\csname url\endcsname\relax
  \def\url#1{\texttt{#1}}\fi
\expandafter\ifx\csname urlprefix\endcsname\relax\def\urlprefix{URL }\fi
\expandafter\ifx\csname href\endcsname\relax
  \def\href#1#2{#2} \def\path#1{#1}\fi

\bibitem{Eisert07}
C.~Dawson, J.~Eisert, T.~J. Osborne, {Unifying Variational Methods for
  Simulating Quantum Many-Body Systems}, Phys. Rev. Lett. 100 (2008) 130501.

\bibitem{ALPS}
{ALPS Consortium}, {The ALPS project (Algorithms and Libraries for Physics
  Simulations) is an open source effort aiming at providing high-end simulation
  codes for strongly correlated quantum mechanical systems as well as $C++$
  libraries for simplifying the development of such code},
  http://alps.comp-phys.org/mediawiki/index.php/Main-Page.

\bibitem{GR06}
J.~J. Garc{\'i}a-Ripoll, {Time Evolution of Matrix Product States}, New J.
  Phys. 8 (2006) 305.

\bibitem{PMCV10}
B.~Pirvu, V.~Murg, J.~I. Cirac, F.~Verstraete,
  \href{http://dx.doi.org/10.1088/1367-2630/12/2/025012}{{Matrix Product
  Operator Representations}}, New J. Phys. 12 (2010) 025012.
\newline\urlprefix\url{http://dx.doi.org/10.1088/1367-2630/12/2/025012}

\bibitem{VCM09}
F.~Verstraete, I.~Cirac, V.~Murg, {Matrix Product States, Projected Entangled
  Pair States, and Variational Renormalization Group Methods for Quantum Spin
  Systems}, Adv. Phys. 57 (2008) 143--224.

\bibitem{nakahara2008quantum}
M.~Nakahara, T.~Ohmi, {Quantum Computing: From Linear Algebra to Physical
  Realizations}, CRC Press, 2008.

\bibitem{P70}
P.~Pfeuty, {The One-Dimensional Ising Model with a Transverse Field}, Ann.
  Phys. 57 (1970) 79--90.

\bibitem{LSM61}
E.~Lieb, T.~Schultz, D.~Mattis, {Two Soluble Models of an Antiferromagnetic
  Chain}, Ann. Phys. 16 (1961) 407--466.

\bibitem{AKLT87}
I.~Affleck, T.~Kennedy, E.~H. Lieb, H.~Tasaki, {Rigorous Results on
  Valence-Bond Ground States in Antiferromagnets}, Phys. Rev. Lett. 59 (1987)
  799--802.

\bibitem{Davis94Circulant}
P.~J. Davis, {Circulant Matrices}, AMS Chelsea Publishing, 1994.

\bibitem{Tyrtyshnikov00Circulant}
S.~S. Capizzano, E.~Tyrtyshnikov, {Any Circulant-Like Preconditioner for
  Multilevel Matrices Is Not Superlinear}, SIAM J. Matrix Anal. Appl. 21 (2000)
  431--439.

\bibitem{CantoniButler}
A.~Cantoni, P.~Butler, {Properties of the Eigenvectors of Persymmetric Matrices
  with Applications to Communication Theory}, IEEE Trans. Commun. 24 (1976)
  804--809.

\bibitem{Plenio05}
M.~Plenio, J.~Eisert, J.~Dreissig, M.~Cramer, {Entropy, Entanglement, and Area:
  Analytical Results for Harmonic Lattice Systems}, Phys. Rev. Lett. 94 (2005)
  060503.

\bibitem{Plenio06}
M.~Cramer, J.~Eisert, M.~Plenio, J.~Dreissig, {Entanglement-Area Law for
  General Bosonic Harmonic Lattice Systems}, Phys. Rev. A 73 (2006) 012309.

\bibitem{Wolf07b}
M.~Wolf, F.~Verstraete, M.~B. Hastings, I.~Cirac, {Area Laws in Quantum
  Systems: Mutual Information and Correlations}, Phys. Rev. Lett. 100 (2008)
  070502.

\bibitem{ECP10}
J.~Eisert, M.~Cramer, M.~B. Plenio, {Area Laws for the Entanglement Entropy},
  Rev. Mod. Phys. 82 (2010) 277--306.

\bibitem{Hastings07}
M.~B. Hastings, {An Area Law for One-Dimensional Quantum Systems}, J. Stat.
  Mech. Theor. Exp. 2007 (2007) P08024.

\bibitem{Hastings07b}
M.~B. Hastings, {Entropy and Entanglement in Quantum Ground States}, Phys. Rev.
  B 76 (2007) 035144.

\bibitem{VC06}
F.~Verstraete, J.~I. Cirac, {Matrix Product States Represent Ground States
  Faithfully}, Phys. Rev. B 73 (2006) 094423.

\bibitem{Schuch09}
N.~Schuch, I.~Cirac, {Matrix Product State and Mean Field Solutions for
  One-Dimensional Systems can be Found Efficiently}, Phys. Rev. A 82 (2010)
  012314.

\bibitem{Fannes92a}
M.~Fannes, B.~Nachtergaele, R.~Werner, {Abundance of Translation Invariant Pure
  States on Quantum Spin Chains}, Lett. Math. Phys. 25 (1992) 249--258.

\bibitem{Fannes92b}
M.~Fannes, B.~Nachtergaele, R.~F. Werner, {Finitely Correlated States on
  Quantum Spin Chains}, Commun. Math. Phys. 144 (1992) 443--490.

\bibitem{LNP528}
I.~Peschel, X.~Wang, M.~Kaulke, K.~Hallberg (Eds.), {Density-Matrix
  Renormalization: A New Numerical Method in Physics}, Lecture Notes in Physics
  Vol.~528, Springer, Berlin, 1999.

\bibitem{Schollwoeck05}
U.~Schollw{\"o}ck, {The Density-Matrix Renormalization Group}, Rev. Mod. Phys.
  77 (2005) 259--315.

\bibitem{VC04a}
F.~Verstraete, J.~I. Cirac, {Valence-Bond States for Quantum Computation},
  Phys. Rev. A 70 (2004) 060302.

\bibitem{VC04b}
F.~Verstraete, J.~I. Cirac, {Renormalization Algorithms for Quantum-Many Body
  Systems in Two and Higher Dimensions}, http://arXiv.org:cond-mat/0407066
  (2004).

\bibitem{Anders06}
S.~Anders, M.~B. Plenio, W.~D{\"u}r, F.~Verstraete, H.~J. Briegel,
  {Ground-State Approximation for Strongly Interacting Spin Systems in
  Arbitrary Spatial Dimension}, Phys. Rev. Lett. 97 (2006) 107206.

\bibitem{Vidal07}
G.~Vidal, {Entanglement Renormalization}, Phys. Rev. Lett. 99 (2007) 220405.

\bibitem{Schuch08a}
N.~Schuch, M.~Wolf, F.~Verstraete, I.~Cirac, {Strings, Projected Entangled Pair
  States, and Variational Monte Carlo Methods}, Phys. Rev. Lett. 100 (2008)
  040501.

\bibitem{Eisert08}
R.~H{\"u}bner, C.~Kruszynska, L.~Hartmann, W.~D{\"u}r, F.~Verstraete,
  J.~Eisert, M.~Plenio, {Renormalisation Algorithm with Graph Enhancement},
  Phys. Rev. A 79 (2009) 022317.

\bibitem{HJ1}
R.~Horn, C.~Johnson, Matrix Analysis, Cambridge University Press, Cambridge,
  1987.

\bibitem{CarrollChang70ALS}
J.~Carroll, J.~J. Chang, {Analysis of Individual Differences in
  Multidimensional Scaling via an $n$-Way Generalization of Eckart-Young
  Decomposition}, Psychometrika 35 (1970) 283--319.

\bibitem{Multi-way}
A.~Smilde, R.~Bro, P.~Geladi, {Multi-Way Analysis. Applications in the Chemical
  Sciences}, Wiley, 2004.

\bibitem{Affleck85}
I.~Affleck, {Large-$n$ Limit of $SU(n)$ Quantum "Spin" Chains}, Phys. Rev.
  Lett. 54 (1985) 966--969.

\bibitem{Vidal03}
G.~Vidal, {Efficient Classical Simulation of Slightly Entangled Quantum
  Computations}, Phys. Rev. Lett. 91 (2003) 147902.

\bibitem{Delgado01}
M.~A. Mart\protect{\'\i{}}n-Delgado, M.~Roncaglia, G.~Sierra, {Stripe
  Ans{\"a}tze from Exactly Solved Models}, Phys. Rev. B 64 (2001) 075117.

\bibitem{Wilson75}
K.~G. Wilson, {The Renormalization Group: Critical Phenomena and the Kondo
  Problem}, Rev. Mod. Phys. 47 (1975) 773--840.

\bibitem{White92}
S.~R. White, {Density Matrix Formulation for Quantum Renormalization Groups},
  Phys. Rev. Lett. 69 (1992) 2863--2866.

\bibitem{Schuch08b}
N.~Schuch, I.~Cirac, F.~Verstraete, {Computational Difficulty of Finding Matrix
  Product Ground States}, Phys. Rev. Lett. 100 (2008) 250501.

\bibitem{PerezGarcia07}
D.~P{\'e}rez-Garc{\'i}a, F.~Verstraete, M.~M. Wolf, J.~I. Cirac,
  {Matrix-Product State Representations}, Quant. Inf. Comput. 7 (2007)
  401--430.

\bibitem{SchollDMRG2011}
U.~Schollw\"ock, {The Density-Matrix Renormalization Group in the Age of Matrix
  Product States}, Ann. Phys. 326 (2011) 96--192.

\bibitem{Eckholt11Matrix}
M.~G. Eckholt-Perotti, Matrix product formalism, Master's thesis, Technische
  Universit\"at M\"unchen and Max-Planck-Institut f\"ur Quantenoptik (2005).

\bibitem{Huckle11Exploiting}
T.~Huckle, K.~Waldherr, T.~Schulte-Herbr{\"u}ggen, {Exploiting Matrix
  Symmetries and Physical Symmetries in Matrix Product States and Tensor
  Trains}, Lin. Multilin. Alg. 61 (2013) 91--122.

\bibitem{Oseledets11tt}
I.~V. Oseledets, {Tensor-Train Decomposition}, SIAM J. Sci. Comput. 33 (2011)
  2295--2317.

\bibitem{Verstraete04a}
F.~Verstraete, D.~Porras, I.~Cirac, {DMRG and Periodic Boundary Conditions: A
  Quantum Information Perspective}, Phys. Rev. Lett. 93 (2004) 227205.

\bibitem{MPDO2004}
F.~Verstraete, J.~J. Garc\'\i{}a-Ripoll, J.~I. Cirac, {Matrix Product Density
  Operators: Simulation of Finite-Temperature and Dissipative Systems}, Phys.
  Rev. Lett. 93 (2004) 207204.

\bibitem{Oseledets10ttm}
I.~V. Oseledets, {Approximation of $2^d \times 2^d$ matrices using tensor
  decomposition}, SIAM J. Matrix Anal. Appl. 31 (2010) 2130--2145.

\bibitem{VWPGC06}
F.~Verstraete, M.~M. Wolf, D.~P{\'e}rez-Gar{\'i}a, I.~Cirac, {Criticality, the
  Area Law, and the Computational Power of Projected Entangled Pair States},
  Phys. Rev. Lett. 96 (2006) 220601.

\bibitem{Schuch07}
N.~Schuch, M.~M. Wolf, F.~Verstraete, I.~Cirac, {Computational Complexity of
  Projected Entangled Pair States}, Phys. Rev. Lett. 98 (2007) 140506.

\bibitem{PEPS2007}
M.~Aguado, J.~I. Cirac, G.~Vidal, {Topology in Quantum States. PEPS Formalism
  and beyond}, J. Phys. Conf. Ser. 87 (2007) 012003.

\bibitem{VC10}
F.~Verstraete, J.~I. Cirac, {Renormalization and Tensor Product States in Spin
  Chains and Lattices}, Phys. Rev. A 42 (2010) 504004.

\bibitem{TensorReview}
T.~G. Kolda, B.~W. Bader, {Tensor Decompositions and Applications}, SIAM Rev.
  51 (2009) 455--500.

\bibitem{bestrank-1}
L.~de~Lathauwer, B.~de~Moor, J.~Vandewalle, {On the Best Rank-1 and
  Rank-($R_1,R_2,\dots,R_n$) Approximation of Higher-Order Tensors}, SIAM J.
  Matrix Anal. Appl. 21 (2000) 1324--1342.

\bibitem{linalgtensor09}
I.~V. Oseledets, D.~V. Savostyanov, E.~E. Tyrtyshnikov, {Linear Algebra for
  Tensor Problems}, Computing 85 (2009) 169--188.

\bibitem{tree-tucker}
I.~V. Oseledets, E.~E. Tyrtyshnikov, {Breaking the Curse of Dimensionality, or
  How to Use SVD in Many Dimensions}, SIAM J. Sci. Comput. 31 (2009)
  3744--3759.

\bibitem{OseTyr09Recursive}
I.~V. Oseledets, E.~E. Tyrtyshnikov, {Recursive Decomposition of
  Multidimensional Tensors}, Dokl. Math. 80 (2009) 460--462.

\bibitem{Khoromskij11ApproximationTC}
B.~Khoromskij, {$\mathcal O(d \log N)$-Quantics Approximation of $N-d$ Tensors
  in High-Dimensional Numerical Modeling}, Constructive Approximation 34 (2011)
  1--24.

\bibitem{Verstraete06}
{F.~Verstraete and J.~I.~Cirac}, {Matrix Product States Represent Ground States
  Faithfully}, Phys. Rev. B 73 (2006) 094423.

\bibitem{ELS08}
M.~Rajih, P.~Comon, R.~A. Harshman, {Enhanced Line Search: A Novel Method to
  Accelerate Parafac}, SIAM J. Matrix Anal. Appl. 30 (2008) 1128--1147.

\end{thebibliography}

\end{document}